\documentclass[11pt]{article}%
\usepackage{makeidx}
\usepackage{amssymb}
\usepackage[all]{xy}
\usepackage[latin1]{inputenc}
\usepackage{amsfonts}
\usepackage{amsmath}
\usepackage{amssymb}

\usepackage{url}
\usepackage{hyperref}
\usepackage{graphicx}%
\usepackage{amsthm}
\usepackage{url}
\usepackage{hyperref}
\providecommand{\U}[1]{\protect\rule{.1in}{.1in}}

\usepackage{xcolor}

\newtheorem{theorem}{Theorem}[section]

\newtheorem{prop}{Proposition}[section]

\newcounter{hypA}

\makeatother

\newcommand{\EE}{\mathbb{E}}

\newcommand{\LL}{\mathbb{L}}

\newcommand{\NN}{\mathbb{N}}
\newcommand{\PP}{\mathbb{P}}

\newcommand{\RR}{\mathbb{R}}

\newcommand{\ZZ}{\mathbb{Z}}

\newcommand{\Ka}{ {\cal K }}

\newcommand{\Sa}{ {\cal S}}

\newcommand{\Za}{ {\cal Z }}

\newcommand{\Xb}{ {\bf X }}

\newcommand{\point}{\mbox{\LARGE .}}

\newcommand{\cqfd}{\hfill\blbx \\}
\def\blbx{\hbox{\vrule height 5pt width 5pt depth 0pt}\medskip}

\def \PP{\mathbb{P}}
\def \RR{\mathbb{R}}
\def \EE{\mathbb{E}}

\def \LL{\mathbb{L}}

\def \ZZ{\mathbb{Z}}
\def \SS{\mathbb{S}}

\def \QQ{\mathbb{Q}}

\usepackage{tikz}
\usepackage{authblk}

\def \chir{\zeta}

\usepackage{bbm}

\usepackage{graphicx}

\begin{document}

  \title{A Note on Random Walks with Absorbing barriers and Sequential Monte Carlo Methods}
\author[1]{Pierre Del Moral\thanks{p.del-moral@unsw.edu.au}}
\author[2]{Ajay Jasra\thanks{staja@nus.edu.sg}}
\affil[1]{INRIA
Bordeaux Sud-Ouest Research Center, FR \& School Mathematics and Statistics, University of New South Wales, AUS}
\affil[2]{Department of Statistics  Applied Probability, National University of Singapore, SG}


\maketitle

\begin{abstract}
In this article we consider importance sampling (IS) and sequential Monte Carlo (SMC) methods
in the context of $1$-dimensional random walks with absorbing barriers. In particular,
we develop a very precise variance analysis for several IS and SMC procedures.
We take advantage of some explicit spectral formulae available for these models to derive
sharp and explicit estimates; this provides stability properties of the associated
normalized Feynman-Kac semigroups. 
Our analysis allows one to compare the variance of SMC and IS techniques for these models. 
The work in this article, is one of the few to consider an in-depth analysis of an SMC method for a particular model-type
as well as variance comparison of SMC algorithms.\\

\end{abstract}


\section{Introduction}

Random walks with absorbing barriers are the simplest example of stochastic processes confined in some domain. In the simplest form, a particle starts
at the origin of the straight line and move back and forth as a simple random walk. When the total displacement is larger than some given critical
value, the particle gets absorbed. The problem is to determine the statistics of the absorption time, as well as the conditional
distributions of the particle  before absorption at any time and their limits when the time horizon tends to $\infty$. These limiting distributions
are often called quasi-invariant distributions or Yaglom measures in reference to one of the pioneering work of Yaglom on this subject in the context of branching processes~\cite{yaglom}.

The origin of these stochastic models certainly started with the Gambler's ruin problem proposed by Pascal to Fermat in 1656, also known as Huygens fifth problem. The seminal book of Feller provides a modern description of these problems and their applications in game 
theory~\cite{feller}.
Conditional Markov chains of this type are also used in biology and branching processes~\cite{ferrari},
as well as in molecular physics~\cite{novikov}, medicine~\cite{bell}, queuing theory~\cite{bohm},
signal processing and rare even analysis~\cite{delmoral1,delmoral2,dmiclo}.

The distribution flow of the random states
of these confined Markov chains satisfy a filtering type nonlinear equation in the space of probability measures on the state space. These equations are decomposed into two sets: The first one is a Bayes' type formula. It transforms the distribution of an internal state of the chain by its conditional distribution 
restricted to the domain of interest. The second one is a simple Markov transport equation. It represents the evolution of the states before
a possible absorption. In this interpretation, a quasi-invariant distribution is a fixed point of these nonlinear equations. For a more thorough discussion on these limiting distributions in terms of Feynman-Kac semigroups we refer the reader to~\cite{dg1,dg2,dmiclo}, and 
the more recent studies \cite{dv,diaconis-miclo}.

Another important problem
is to compute the conditional distributions of a non absorbed trajectory. 
 These complex 
measures are defined in terms of Feynman-Kac distributions on  path-spaces~\cite{dd,dmiclo}. They represent the conditional probability of the historical process of the chain given it has not been absorbed. These path space measures are expressed as the distribution
of the trajectories of a given Markov chain weighted by some product of non negative potential-type functions. These functions represent the non
absorption probabilities at any given state. The numerical integration of Feynman-Kac measures is often performed using Monte Carlo methods.

Several strategies can be developed: The first idea relies on importance sampling techniques. The objective is to find a judicious twisted Markov chain that mimic the evolution of the underlying non absorbed process. One of the main drawback of this technique is that the variance of the importance sampling weights is often degenerate w.r.t.~the time parameter; this is discussed below. Another drawback is that is it an intrusive technique, in the sense that it requires to change the physical evolution of the underlying Markov chain.  Another strategy is to interpret these importance sampling weights as non absorption probability. When a particle is absorbed, instantly another one duplicates. The selection of the duplicated particle is performed randomly with a probability proportional to its weight. For a more thorough discussion on these particle samplers (sequential Monte Carlo methods) and their application domains we refer the reader to the research monographs~\cite{delmoral1,delmoral2,dmiclo}, the more recent studies~\cite{chan2} and the references therein. The convergence analysis of these
particle samplers as the number of particles tends to $\infty$ has been developed in various directions, including propagation of chaos, fluctuation theorems and large deviation principle. The analysis of the long time behavior of these samplers have been started in~\cite{dg1,dg2,dmiclo,dmiclo2}, including uniform variance and exponential concentration estimates w.r.t. the time horizon.

Nevertheless most of the stochastic analysis developed in these works is expressed in terms of mathematical objects
which are difficult to quantify as their rely on the stability properties of normalized non absorption semigroups. As a result it is rather difficult to compare the  variance of
particle sampler with more conventional Monte Carlo methods based on importance sampling techniques; this is the main topic of this article, for absorbing random walks.

The comparison of variances or efficiency in estimation of different algorithms, for the same quantity, is 
an important issue in Monte Carlo approximation. For instance, this is considered in many
works such as \cite{chen,giles_acta} and in particular for SMC algorithms in \cite{chopin,dj,whiteley}.
Comparing SMC algorithms is not a trivial task and as shown by the relatively few articles on this subject.
For instance \cite{chopin} compares SMC algorithms which resample multinomially and with the
residual method, showing that the asymptotic variance in the CLT of the latter is smaller than the former.
\cite{dj} also consider variance comparison, but for linear Gaussian models. Perhaps the most complete
approach is in \cite{whiteley}, where the fixed $N$, but asymptotic in $n$, (the time parameter) properties
of second moments of  normalizing constant quantities are considered. Rather substantial work is required to compare algorithms
using the approach of \cite{whiteley}, but several interesting and sometimes known facts are derived. In general,
however, it is not always trivial to compare SMC algorithms and certainly one important way is by considering specific
examples where explicit calculations can be performed.

In this article, as already mentioned, we consider the comparison of limiting variances for the absorbing random walk example,
for several different SMC and IS procedures. We note that for comparison between IS and SMC in estimation,
generally the latter performs substantially better w.r.t.~the time horizon $n$. For instance, in the estimation of `marginal' quantities
(as we explain later) under reasonably weak conditions \cite{delmoral1,delmoral2,whiteley1} the variance of SMC is time-independent and the relative variance of the estimate
of the `normalizing constant' (e.g.~the probability that the time to  absorbtion is bigger than $n$) is linear in $n$ \cite{cdg:11}.
However, often, e.g.~for the normalizing constant for IS the same quantity is exponential in $n$; see \cite{cdg:11} and the discussion
in \cite{whiteley}. However, this need
not be the case for instance if the target and proposal do not become mutually singular in the limit as $n$ grows. This
latter property is not always simple to check and can suprisingly occur in some examples e.g.~\cite{wang}. Therefore,
in our absorbing random walk example, an in-depth analysis is of interest, as \emph{a priori} it is not completely
obvious that SMC outperforms IS. Our work is also of interest, as it is one of the few articles where one considers
analysis of SMC in a very specific example. It is remarked that due to the complexity of the associated calculations,
to get very explicit results, one cannot consider a very challenging model. In our example, one does not need IS
nor SMC as some very efficient Doob-type transformations can be derived and directly sampled. The point here
is to provide an in-depth analysis, which could be of use in more complex models.

The main objective of the article is to improve the understanding of sequential Monte Carlo methods and the stability properties of 
normalized Feynman-Kac semigroups in the context of $1$-dimensional random walks with absorbing barriers. The central idea is to take
 advantage of some explicit spectral formulae available for these models to derive
sharp and explicit estimates. 
Our analysis allows one to compare the variance of particle samplers with the one of importance sampling type techniques based on random walks with repulsive boundaries. 

This article is structured as follows:  In section \ref{sec:notation} the notation for the article is established. In section \ref{description-models-sec} we 
give details of the model under study as well as some properties of them. In section \ref{smc-section} our algorithms as well as our variance results are stated.
In section \ref{sec:fk}, some Feynman-Kac notions are discussed and the proofs of the results in section \ref{description-models-sec} are stated.
Section \ref{var-analysis} gives the proofs of our variance results in section \ref{smc-section}. The appendix contains some calculations that are used at various
places in the article.
 
 \section{Notation}\label{sec:notation}

In the further development of the article $S$ stand of the finite set $S=\{1,\ldots,d\}\subset \NN$, for some $d$, equipped with the uniform counting 
measure $x\in S\mapsto u(x)=1/d$. The boundary of the set $S$ is the set $\partial S=\{0,d+1\}$.
We use as a rule the bold letters  $\boldsymbol{X_n}:=(X_p)_{0\leq p\leq n}\in S_n:=S^{n+1}$
to denote the historical process associated with a given Markov chain $X_n$ on $S$. 
We also use the lower index 
$f_n$ to denote functions on the path spaces $S_n$. The total variation distance between two probability measures
$\mu_1$ and $\mu_2$ on $S$ is defined by
$$
\Vert \mu_1-\mu_2\Vert_{\tiny tv}:=\frac{1}{2}~\sum_{x\in S}\vert  \mu_1(x)-\mu_2(x)\vert.
$$
For any $x\in S$, we let $y\in S\mapsto \delta_x(y)=1_{x=y}$ the Dirac measure  at the state $x$.
$$
\beta(M):=\sup_{(x,y)\in S^2}\Vert\delta_xM-\delta_yM\Vert_{\tiny tv}.
$$
Given for positive function $f$ on $S$ we set
$
\rho(f):=\sup_{(x,y)\in S^2}{[f(x)/f(y)]}.
$.
 The Boltzmann-Gibbs transformation $\Psi_{g}$ from the set of probability measures on $S$ into itself is defined for any
 probability measure $\mu$ on S by the probability measure 
$$
x\in S\mapsto \Psi_{g}(\mu)(x):=\frac{1}{\mu(g)}~g(x)~\mu(x)\quad\mbox{\rm with}\quad
\mu(g):=\sum_{x\in S}\mu(x)g(x).
$$
Given some probability measure $\mu$ on $S$ we let $\LL_2(\mu)$ be the Hilbert space equipped with the 
inner product and the norm
$$
\langle f_1,f_2\rangle_{\mu}=\mu(f_1f_2)\quad\mbox{\rm and}\quad
\Vert f\Vert_{2,\mu}^2=\mu(f^2).
$$

\section{Description of the models}\label{description-models-sec}
\subsection{A random walk with absorbing barriers}
Let $K$ be the transition probabilities of a symmetric random walk $X_n$ on the $1$-dimensional lattice $\ZZ$ given by
$$
K(i,i-1)=K(i,i+1)=\frac{1}{2+\theta}\quad\mbox{\rm and}\quad K(i,i)=\frac{\theta}{\theta+2}
$$
where $\theta$ is a given non-negative parameter. We write $K_0$ the transition of the simple random walk associated with 
the null parameter $\theta=0$.
We assume that $X_0$ has some probability
distribution $\eta_0$  on $S$. 

Let $T_X$ be the first time the chain $X_n$ exits the set
$S$.  We are interested in computing the quantities
\begin{equation}\label{def-eta-Q-Za}
\eta_{n}:=\mbox{\rm Law}(X_n~|~T_X>n)\qquad \Za_n:=\PP(T_X>n)\quad\mbox{\rm and}\quad
\QQ_{n}:=\mbox{\rm Law}(\boldsymbol{X_n}~|~T_X>n).
\end{equation}

 The stochastic model discussed above 
can be interpreted as an absorbed random walk $X^c_n$ killed at the barriers $\partial S$. 
The killing probabilities are defined by
the indicator functions $1-1_{S}$. When the chain hits one of the the barriers $\partial S$ it goes to a cemetery (a.k.a.~coffin) state denoted by 
the letter $c$.
The killing time of the chain evolving in the augmented state space $S\cup\{c\}$ coincides with $T_X$. 

 If $\widehat{\eta}_{n+1}:=\eta_{n}K$
($\Rightarrow \eta_n= \Psi_{1_S}\left(\widehat{\eta}_n\right)$)
stands for the conditional distribution of the random state $X_{n+1}$ given $T_X>n$, then we have the geometric formula
\begin{equation}\label{geometric-formula}
\PP(T_X=n+1)=\left\{\prod_{1\leq p\leq  n}\widehat{\eta}_{p}(1_S)\right\}~\left(1-\widehat{\eta}_{n+1}(1_S)\right)=
\widehat{\gamma}_{n+1}(1-1_S)
\end{equation}
 with the unnormalized Feynman-Kac measures $\widehat{\gamma}_n$ on $S$ defined by
 \begin{equation}\label{geometric-formula-hard}
\widehat{\gamma}_{n+1}(f):=\EE(f(X_{n+1})~1_{T_X}>n)=\widehat{\eta}_{n+1}(f)~
\prod_{1\leq p\leq  n}\widehat{\eta}_{p}(1_S).
 \end{equation}
 Observe that $\widehat{\gamma}_{n}$ satisfies the matrix evolution equation
 $$
 \widehat{\gamma}_{n+1}=\widehat{\gamma}_{n}\widehat{Q}\quad \mbox{\rm with}\quad \widehat{Q}(x,y)=1_S(x)~K(x,y).
 $$
 
The confinement model discussed above can be extended to more general processes evolving in absorbing environments,
including soft obstacles models associated with partially absorbing media. 

The random walk
model with absorbing barriers can be solved using spectral decompositions of symmetric tridiagonal matrices. 
For instance if we consider the symmetric matrix with positive entries
$$
\forall x\in S\qquad Q(x,y):=K(x,y)1_{S}(y)
\quad\mbox{
then we have}\quad
\quad
Q(\varphi_0)=E_0~\varphi_0
$$
with the top eigenvalue of $Q$ given by
$$
E_0
:=1-\frac{2}{1+\theta/2}~\sin^2{\left(\frac{\pi}{2(d+1)}\right)}~\in~ ]0,1[
$$
and the corresponding $\LL_2(u)$-normalized eigenfunction
$$
\varphi_0(x):=\sqrt{\frac{2d}{d+1}}
\sin{\left(\frac{x\pi}{d+1}\right)}.
$$
A more refined spectral analysis yields the following theorem.
\begin{theorem}
For any starting states $(x,y)\in S^2$ and any time horizon $n$ we have
\begin{equation}\label{Qn1-rho-ref}
\PP(T_X>n~|~X_0=x)\leq \frac{2+\theta}{1+\theta}~\sin^{-1}{\left(\frac{\pi}{d+1}\right)}~~\PP(T_X>n~|~X_0=y).
\end{equation}
In addition, we have
\begin{equation}\label{Qn1-ref}
\sin{\left(\frac{\pi}{d+1}\right)} \leq E_0^{-n}~\PP(T_X>n~|~X_0=x)\leq \sin^{-1}{\left(\frac{\pi}{d+1}\right)}.
\end{equation}
\end{theorem}
The proof of the theorem is provided in section~\ref{some-quantitative-sec}.

The confinement distribution $\QQ_n$ and their normalizing constants $\Za_n$ can be expressed in terms of the Doob
 $\varphi_0$-process.
This process is defined by the Markov chain $Y^{\varphi}_n$ on $S$ with initial distribution 
$
\displaystyle\eta^{\varphi}_0(x):=\Psi_{\varphi_0}(\eta_0)(x)
$
and elementary Markov transitions
$$
M_{\varphi}(x,y):=\frac{Q(x,y)~\varphi_0(y)}{Q(\varphi_0)(x)}.
$$
In this situation, we have
\begin{equation}\label{ref-norm-H-E0-intro}
\EE\left(\boldsymbol{f_n(X_n)}~|~T_X>n\right)~\propto ~\EE\left(\boldsymbol{f_n(Y^{\varphi}_n)}/\varphi_0(Y^{\varphi}_n)\right)\quad\mbox{\rm and}\quad
\Za_n=E_0^{n}~\eta_{0}(\varphi_0)~\EE(1/\varphi_0(Y^{\varphi}_n)).
\end{equation}
 We also have the limiting quasi-invariant distributions
 $$
 \displaystyle\lim_{n\rightarrow\infty} \eta_n(x)=\pi(x):=\Psi_{\varphi_0}(u)(x)=\displaystyle\tan{\left(\frac{\pi}{2(d+1)}\right)}~\sin{\left(\frac{x~\pi}{d+1}\right)}
 $$
 and
 $$
\displaystyle\lim_{n\rightarrow\infty} \PP(Y^{\varphi}_n=x)=\pi_{\varphi}(x):=\Psi_{\varphi_0}(\pi)(x)=\frac{2}{d+1}~
\sin^2{\left(\frac{x\pi}{d+1}\right)}.
$$

All of these formulae can be used to compute explicitly the quantities (\ref{def-eta-Q-Za}) and the limiting quasi-invariant
distributions without to resort to any Monte Carlo approximation. Conversely, we can use these explicit descriptions
to improve the understanding of sequential Monte Carlo samplers of conditional Markov chains models. Our first step in this direction is to 
quantify the stability properties of the nonlinear semigroups of conditional Markov chains.

The stability properties of the flow of probability measures $\eta_n$ introduced in  (\ref{def-eta-Q-Za}) 
are expressed in terms of the following quantities
\begin{eqnarray*}
\overline{E}_1
&=&1-4~\sin{\left(\frac{3}{2}~\frac{\pi}{d+1}\right)}\sin{\left(\frac{1}{2}~\frac{\pi}{d+1}\right)}~\left(\theta+2\cos{\left(\frac{\pi}{d+1}\right)}\right)^{-1}
\end{eqnarray*}
and the parameters
$$
s_k(d)=\sqrt{\frac{d+1}{2}}~\sin^{-k}{\left(\frac{\pi}{d+1}\right)}\quad\mbox{\rm with}\quad k\in\{1,2,3\}.
$$
We also let $\eta_n^{\prime}$ the distributions defined as $\eta_n$ by replacing $\eta_0$ by some possibly different initial
distribution  $\eta_0^{\prime}$. In this notation, our main result takes basically the following form.
\begin{theorem}\label{stab-theorem}
For any function $f$ s.t. $\pi_{\varphi}(f)=0$ and any $n\geq 0$ we have
\begin{equation}\label{L2+beta}
\Vert M_{\varphi}^n(f)\Vert_{2,\pi_{\varphi}}\leq  \overline{E}_1^{\,n}~
~\Vert f\Vert_{2,\pi_{\varphi}}\quad\mbox{and}\quad
\beta\left(M_{\varphi}^n\right)\leq ~s_1(d)~\overline{E}_1^{\,n}.
\end{equation}
In addition, we have
\begin{equation}\label{contract+pi}
\Vert \eta_{n}-\pi\Vert_{\tiny tv}\leq~s_2(d)~\overline{E}_1^{\,n}~\quad\mbox{and}\quad
\Vert \eta_{n}-\eta_n^{\prime}\Vert_{\tiny tv}
\leq ~s_3(d)~~\frac{2+\theta}{1+\theta}~~\overline{E}_1^{\,n}~~
\Vert  \eta_{0}-\eta_0^{\prime}\Vert_{\tiny tv}.
\end{equation}
\end{theorem}
The proof of the l.h.s. of (\ref{L2+beta}) relies on spectral decompositions of the Markov transitions of the $\varphi_0$-process and it
is provided in section~\ref{spectral-decomposition-sec}. The r.h.s. of (\ref{L2+beta}) is proved in section~\ref{proposition-ratio-functions}.

The power matrices $(Q^n,\widehat{Q}^n)$ and the conditional distributions $(\eta_n,\widehat{\eta}_n)$ can be interpreted in terms of Feynman-Kac semigroups. 

The stability properties of these semigroups have been developed in~\cite{delmoral1,dg1,dg2,dmiclo,dmiclo2,dv,diaconis-miclo}, including for continuous time absorption type process on abstract measurable spaces with soft and hard obstacles. Section~\ref{review-FK-sec} provides a brief review on these semigroups and their Lipschitz regularity properties.

These Feynman-Kac models also allow to interpreted Sequential Monte Carlo samplers and genealogical tree based samplers 
as  mean field particle
interpretation of nonlinear semigroups. The variances of these Monte Carlo schemes are also expressed in terms of Feynman-Kac semigroups.
For the convenience of the reader a description of these probabilistic models is provided in the appendix
on page~\pageref{mean-field-section}. To get more explicit descriptions of the estimates presented in this article w.r.t. the parameter $d$ the  last section of the appendix provide elementary second order expansions of trigonometric
functions and their inverse.

\subsection{A soft obstacle model with repulsive barriers}

Sampling the distributions $\QQ_n$ using acceptance conventional rejection techniques amounts of sampling independent copies of the chain $X_n$, the ones being absorbed are rejected.
The relative variance of these rather crude Monte Carlo estimators is inversely proportional to the non absorption time probabilities. For any 
finite absorbing barriers these relative variance estimates increase exponentially fast w.r.t. the time parameter. 

Another natural idea presented in~\cite{dd} is to turn the hard obstacle into a soft one.
These soft obstacle models are based on the following observation 
\begin{equation}\label{def-Q-theta}
\forall x,y\in S\qquad
K(x,y)~1_{S}(y)=g(x)~M(x,y):=Q(x,y)
\end{equation}
with the function $g$ and the stochastic matrix $M$ on $S$ defined by 
 for any $1<x<d$, 
\begin{eqnarray*}
g(x)&=&1\\
M(x,x-1)&=&K(x,x-1)=M(x,x+1)=K(x,x+1)\quad\mbox{\rm and}\quad M(x,x)=K(x,x)
\end{eqnarray*}
with the boundary conditions
\begin{eqnarray*}
g(1)&=&g(d)=\frac{1+\theta}{2+\theta}\\
M(1,1)&=&1-M(1,2)=\frac{\theta}{1+\theta}=M(d,d)=1-M(d,d-1).
\end{eqnarray*}

We let  $Y_n$ be a Markov chain on  $S$, with Markov transitions $M$ and
initial distribution $\eta_{0}$. Observe that the chain $Y_n$ is reflected at the boundaries of the set $S$, and the importance sampling
 function $g(x)$ represents the chance to stay in the set $S$ starting from the state $x$.

If $\PP_{n}:=\mbox{\rm Law}(\boldsymbol{Y_n})$ then we have the Feynman-Kac formula
  \begin{equation}\label{def-Qa}
 d\QQ_{n}=\Za_n^{-1}~Z_n(Y)~d\PP_{n}\quad\mbox{\rm with}\quad
 Z_n(Y):=\prod_{0\leq p<n}~g(Y_p) 
\quad\mbox{\rm and}\quad  \Za_{n}:=\EE\left(Z_n(Y)\right).
 \end{equation}
 
 This model can also be interpreted as an absorbed random walk $Y^c_n$ with  killing probabilities are defined by
the indicator functions $1-g$. The chain $Y^c_n$ evolves in two steps: between the killing transitions the chain evolves as the chain $Y_n$.
During the killing transition the chain $Y^c_n=y$ at some state $y\in S$ is killed with probability $1-g(y)$. 
As before,  the chain goes to a cemetery (a.k.a. coffin) state denoted by 
the letter $c$.
 If $T_Y$ stands for  the killing time of the chain, we have
\begin{equation}\label{def-eta-Q-Za-soft}
\eta_{n}:=\mbox{\rm Law}(Y^c_n~|~T_Y\geq n)\qquad \Za_n:=\PP(T_Y\geq n)\quad\mbox{\rm and}\quad
\QQ_{n}:=\mbox{\rm Law}(\boldsymbol{Y^c_n}~|~T_Y\geq n).
\end{equation}
In this situation, the geometric formula (\ref{geometric-formula}) takes the form
$$
\PP(T_Y= n)=\left\{\prod_{0\leq p<n}\eta_p(g)\right\}~\left(1-\eta_n(g)\right)=\gamma_n(1-g)
$$
 with the unnormalized Feynman-Kac measures $\gamma_n$ on $S$ defined by
 \begin{equation}\label{geometric-formula-gamma}
 \gamma_n(f):=\EE(f(Y_n)~Z_n(Y))=\eta_n(f)~\prod_{0\leq p<n}\eta_p(g)~\left(~\Longleftrightarrow~\widehat{\gamma}_{n+1}= \gamma_nK\right).
 \end{equation}
Note that $\gamma_{n}$ satisfies the matrix evolution equation
 \begin{equation}\label{def-Phi}
\gamma_{n+1}={\gamma}_{n}{Q}\Longrightarrow \eta_{n+1}=\Phi\left(\eta_n\right):=\Psi_{g}\left(\eta_n\right)M.
 \end{equation}
 In addition we have
  \begin{equation}\label{def-Pn-ineq}
 \sup_{\mu_1,\mu_2}\Vert \Phi^n\left(\mu_1\right)-\Phi^n\left(\mu_2\right)\Vert_{\tiny tv}=
 \sup_{x_1,x_2}\Vert \Phi^n\left(\delta_{x_1}\right)-\Phi^n\left(\delta_{x_2}\right)\Vert_{\tiny tv}=\beta(P_n)
 \end{equation}
 with the Markov operator $P_n$ defined by
 \begin{equation}\label{def-Pn}
 P_n(f)(x)=\EE\left(f(X_n)~|~T_X> n,~X_0=x\right).
 \end{equation}
 The collection of operators $P_n$ do not form a semigroup. Nevertheless we have the following stability properties.
\begin{theorem}\label{theo-stab-Pn}
For any time horizon $n\geq 0$ have
\begin{equation}\label{betaPn}
\beta(P_n)
\leq ~s_2(d)~\overline{E}_1^{\,n}.~
\end{equation}
In addition we have the contraction inequality
\begin{eqnarray}
\beta(P_{n+\varsigma_{P}})&\leq& e^{-1}~\beta(P_{n})
\qquad\mbox{ and}\qquad \sum_{n\geq 0}\beta(P_{n})^2\leq \frac{1}{1-e^{-2}}~\varsigma_{P}\label{def-varsigmaP}
\end{eqnarray}
with the relaxation time
\begin{eqnarray*}
\varsigma_{P}&:=&{\left[1+\log{\left(\frac{2+\theta}{1+\theta}~s_3(d)
\right)}\right]}\,/\,{
\log{\left(1+\left(\frac{1-\overline{E}_1}{\overline{E}_1}\right)\right)}}\\
&&\\
&&\leq (1+\theta/2)~
\sin^{-2}{\left(\frac{1}{2}~\frac{\pi}{d+1}\right)}
\left[1+\log{\left(\frac{2+\theta}{1+\theta}~s_3(d)\right)}\right]\quad\mbox{ as soon as}\quad d>5.
\end{eqnarray*}
\end{theorem}

The proof of theorem~\ref{theo-stab-Pn}  is provided in section~\ref{some-quantitative-sec}.

\section{Sampling Algoritihms}\label{smc-section}

In the further development of this section the quantities $(\Za_n,\gamma_n,\eta_n,\QQ_{n})$ are approximated by
some random quantities $(\Za_n^{{\tiny a}},\gamma^{{\tiny a}}_n,\eta^{{\tiny a}}_n,\QQ^{{\tiny a}}_{n})$ associated with some
Monte Carlo samplers. The upper index $(\point)^{{\tiny a}}$ represents the class of Monte Carlo technique and we set
\begin{eqnarray*}
v_n^{{\tiny a}}(f)&:=&\lim_{N\rightarrow\infty}N~\EE\left(\left[{\Za_n^{{\tiny a}}}{\Za_n^{-1}}~\eta^{{\tiny a}}_{n}(f)-\eta_n(f)
\right]^2\right)\\
w_n^{{\tiny a}}(f)&:=&\lim_{N\rightarrow\infty}N~\EE\left(\left[\eta_{n}^{{\tiny a}}(f)-\eta_{n}(f)
\right]^2\right).
\end{eqnarray*}
When the Monte Carlo estimates are based on independent samples the above quantities are non asymptotic and we can remove the limit
operation in the above definitions.  The variances $$w_n^{{\tiny\bold a}}(f_n):=
\lim_{N\rightarrow\infty}N~\EE\left(\left[\QQ_{n}^{{\tiny a}}(f_n)-\QQ_{n}(f_n)\right]^2\right)
$$ are defined as $w_n^{{\tiny a}}(f)$ by replacing the sample states
by their historical version, and the test-observable function $f$ on the states by a function $f_n$ on path-space. We also underline that all the Monte Carlo approximation $(\Za_n^{{\tiny a}},\gamma^{{\tiny a}}_n)$ discussed in the article are unbiased.\\

 \subsection{The twisted $\varphi_0$-Doob process}\label{twisted-section-dp}

This importance sampling  scheme  is based on sampling  $N$ independent copies $(Y^{\varphi,i}_n)_{i\geq 1}$ of the Doob $\varphi_0$-process
$Y_n^{\varphi}$. The corresponding Monte Carlo estimates are given by
\begin{eqnarray*}
\Za_n^{{\tiny dp}}&:=&E_0^{n}~\eta_{0}(\varphi_0)~\frac{1}{N}\sum_{1\leq i\leq N}~\varphi_0^{-1}(Y^{\varphi, i}_n)\\
\gamma^{{\tiny dp}}_n&:=&\Za^{{\tiny dp}}_{n}\times \eta^{{\tiny dp}}_{n}
\quad\mbox{\rm with}\quad \eta_{n}^{{\tiny dp}}:=\displaystyle\sum_{1\leq i\leq N}\frac{\varphi_0^{-1}(Y^{\varphi, i}_n)}{\sum_{1\leq j\leq N}\varphi_0^{-1}(Y^{\varphi, j}_n)}~\delta_{Y^{\varphi, i}_n}.
\end{eqnarray*}
The variance of these estimates are given by the formulae
\begin{eqnarray}
w_n^{{\tiny dp}}(f)&=& v_n^{{\tiny dp}}[f-\eta_{n}(f)]\quad\mbox{\rm with}\quad
v_n^{{\tiny dp}}(f)
=~\frac{E_0^n}{\eta_0Q^n(1)}~\eta_0(\varphi_0)~\eta_n(f/\varphi_0)-\eta_n(f)^2.
\label{var-formula-dp}
\end{eqnarray}
The proof of these formula follows elementary variance computations for  independent random variables. For the convenience of the reader a sketch of the proof is provided in the appendix, on page~\pageref{proof-var-formula-dp}. At equilibrium these variances become
$$
\eta_0=\pi\Longrightarrow
v_n^{{\tiny dp}}(f)
=\pi(\varphi_0)~\pi(f/\varphi_0)-\pi(f)^2.
$$

 \subsection{The twisted reflected process}\label{twisted-reflected-sec}

This importance sampling  scheme  is  based on sampling  $N$ independent copies $(Y^i_n)_{i\geq 1}$ of the chain $Y_n$
weighted by the change of probability weights $ Z_n(Y)$. The corresponding Monte Carlo estimates are given by
\begin{eqnarray*}
\Za^{{\tiny IS}}_{n}&:=&\frac{1}{N}\sum_{1\leq i\leq N}Z_n(Y^i)\\
\gamma^{{\tiny IS}}_n&:=&\Za^{{\tiny IS}}_{n}\times \eta^{{\tiny IS}}_{n}\quad\mbox{\rm with}\quad
\eta^{{\tiny IS}}_{n}=~\sum_{1\leq i\leq N}~\frac{Z_n(Y^i)}{\sum_{1\leq j\leq N}Z_n(Y^j)}~\delta_{{Y^i_n}}.
\end{eqnarray*}

The variance of these estimates are expressed in terms of the matrices
\begin{equation}\label{tilde-Q-def}
\widetilde{Q}(x,y)=g^2(x)~M(x,y)\quad\mbox{\rm and the measures}\quad \widetilde{\eta}_n(f)=\eta_0\widetilde{Q}^n(f)/\eta_0\widetilde{Q}^n(1).
\end{equation}
In this notation, we have the formulae
\begin{eqnarray}
w_n^{{\tiny IS}}(f)&=&v_n^{{\tiny IS}}(f-\eta_n(f))\quad\mbox{\rm with}\quad
v_n^{{\tiny IS}}(f)
=\frac{\eta_0\widetilde{Q}^n(1)}{[\eta_0{Q}^n(1)]^2}~
\widetilde{\eta}_n(f^2)-\eta_n(f)^2.
\label{var-formula-is}
\end{eqnarray}

Here again, the proof of these formula follows elementary variance computations for  independent random variables. For the convenience of the reader a sketch of the proof is provided in the appendix, on page~\pageref{proof-var-formula-dp}. By the Frobenius theorem $\widetilde{Q}$ has a positive eigenvalue $\widetilde{E}_0>0$ dominating the norm of the other ones. The next theorem shows that $\widetilde{E}_0$ is much larger than the square of $E_0$. As a consequence the variance of the importance sampler based on twisted reflected random walks
increase exponentially w.r.t. the time horizon. Let
$$
\sigma(\theta)=\frac{\sqrt{1+\theta}}{\sqrt{1+\theta}+\sqrt{2+\theta}}.
$$

\begin{theorem}\label{spectral-comparison-theo}
We have the spectral gap inequalities
$$
\widetilde{E}_0-E_0^{2}\leq  \frac{1}{\left(1+\theta/2\right)^{2}}~\left\{
\sin^2{\left[\frac{\pi}{2(d+1)}\right]}~\left(\theta+2\cos{\left(\frac{\pi}{d+1}\right)}\right)~
+\frac{1}{2}~
\sigma(\theta)
~ \right\}
$$
\begin{eqnarray*}
\widetilde{E}_0-E_0^{2}&\geq &\frac{1}{(1+\theta/2)^2}~\sin^2{\left[\frac{\pi}{2(d+1)}\right]}
\left[\frac{d-1}{d+1}~\theta+2\cos{\left(\frac{\pi}{d+1}\right)} \left\{1+
\frac{2}{d+1}~\sigma(\theta)\right\}\right]
\end{eqnarray*}
In particular we have
\begin{equation}\label{var-tilde-inequality}
c^{-1}
\left( 1+~
\displaystyle\frac{1}{2+\theta}~\frac{4}{3}~\sin^2{\left(\frac{1}{d+1}~\frac{\pi}{2}\right)}\right)^n\leq \frac{\eta_0\widetilde{Q}^n(1)}{[\eta_0{Q}^n(1)]^2}\leq c~({\widetilde{E}_0}/{E_0^{2}})^n
\end{equation}
for some finite constant $c>0$ whose values do not depend on the time horizon.
\end{theorem}

The proof of this theorem is provided in the appendix, on page~\pageref{proof-spectral-comparison-theo}.
We already mention that the estimate (\ref{var-tilde-inequality}) is a consequence of the spectral inequalities stated above (and some elementary estimates on trigonometric functions provided at the end of the proof of the theorem). A proof of this claim is provided in (\ref{tilde-E-Q1}).

 \subsection{Particle sampler with reflection}\label{psr-sec}

This particle scheme is based on sampling $N$ copies $\xi_n:=(\xi^i_n)_{1\leq N}$ 
of the absorbed random walk $Y^c_n$. The transition of the chain $\xi_n\leadsto \xi_{n+1}$ is decomposed into two steps
$\xi_n\leadsto \xi_n^{\tiny killing}\leadsto \xi_{n+1}$. During the killing transition $\xi_n\leadsto \xi_n^{\tiny killing}$
when a particle is absorbed, instantly one of the $N$ particles duplicates. The choice of the duplicated particle, say $\xi^j_n$
is made independently with a probability proportional to its
weight $g(\xi^j_n)$. During the second transition $\xi_n^{\tiny killing}\leadsto \xi_{n+1}$ the particles evolve independently according to
the transition $M$.

The corresponding Monte Carlo estimates are given by

\begin{eqnarray*}
\Za^{{\tiny soft}}_{n}&:=&\prod_{0\leq p< n}~\eta^{{\tiny soft}}_{p}(g)\\
\gamma^{{\tiny soft}}_n&:=&\Za^{{\tiny soft}}_{n}\times \eta^{{\tiny soft}}_{n}\quad\mbox{\rm with}\quad
\eta^{{\tiny soft}}_{n}:=\sum_{1\leq i\leq N}~\delta_{{\xi^i_n}}.
\end{eqnarray*}
The killing transition can be seen as a birth and death transition. 
In this interpretation arise the important notion of the ancestral lines of the particles. In this situation,
the Monte Carlo approximation $\QQ^{{\tiny soft}}_{n}$ of $\QQ_{n}$ is defined as above by replacing ${\xi^i_n}$ by its ancestral line.\\

The variance of these estimates are expressed in terms of  the normalized semigroup
\begin{equation}\label{normalized-fksg}
{Q}_{p,n}:=\frac{{Q}^{n-p}}{\eta_{p}{Q}^{n-p}(1)}~\Longrightarrow~\eta_p{Q}_{p,n}=\eta_n.
\end{equation}
In this notation, we have the formulae
\begin{eqnarray}
\displaystyle w_n^{{\tiny soft}}(f)
&=&v_n^{{\tiny soft}}(f-\eta_n(f))\nonumber\\
\displaystyle v_n^{{\tiny soft}}(f)&=&\sum_{0\leq p\leq   n}\eta_p\left(\left[{Q}_{p,n}(f)-\eta_n(f)\right]^2\right)\nonumber\\
&&\hskip3cm-\sum_{0\leq p<   n}~\eta_{p}(g)^2~
\eta_{p}\left(\left[~{Q}_{p,n}(f)-\frac{g}{\eta_{p}(g)}~\eta_n(f)\right]^2\right) \label{v-soft-f}.
\end{eqnarray}

These fluctuation formulae can be deduced from theorems 14.4.3 and 16.6.2 in~\cite{delmoral2}.
For the convenience of the reader, the details of the proof of (\ref{v-soft-f})  are  provided in the appendix.

The variances $v_n^{{\tiny soft}}(f)$ and $ w_n^{{\tiny\bf soft}}({f_n})$ generally increase linearly w.r.t. the time horizon, while  $w_n^{{\tiny soft}}(f)$ are uniformly bounded w.r.t. the time horizon.
For instance when $\eta_0=\pi$ and $f=\varphi_0$ we have
\begin{equation}\label{v-soft-varphi-ref}
\left((1-E_0)+\frac{1}{n}\right)~\pi\left(\left[\varphi_0-\pi\left(\varphi_0\right)\right]^2\right)\leq 
\displaystyle n^{-1}~v_n^{{\tiny soft}}(\varphi_0)
\leq \left(1+\frac{1}{n}\right)~\pi\left(\left[\varphi_0-\pi\left(\varphi_0\right)\right]^2\right).
\end{equation}

The next proposition provides some useful estimates in terms of the parameters $(d,\theta)$ and the time horizon.

\begin{prop}~\label{estimates-var-soft}
For any function $f$ on $S$ we have
\begin{eqnarray*}
w_n^{{\tiny soft}}(f)
&\leq &\displaystyle  \eta_n\left(\left[f-\eta_n(f)\right]^2\right)+\frac{2}{1+\theta}~\frac{1}{1-e^{-1}}~\sin^{-1}{\left(\frac{\pi}{d+1}\right)}~(\varsigma_{P}-1)~\mbox{\rm osc}(f)\\
 v_n^{{\tiny soft}}(f)&\leq &(n+1)~\frac{2+\theta}{1+\theta}~\left[\sin{\left(\frac{\pi}{d+1}\right)}\right]^{-1}~\Vert f\Vert
 \end{eqnarray*}
with the relaxation time $\varsigma_{P}$ defined in (\ref{def-varsigmaP}).  For any functions ${f_n}$ with at most unit oscillations  on the path spaces $S_n$ we have
\begin{eqnarray*}
 w_n^{{\tiny\bf soft}}({f_n})&\leq& 1+\frac{2n}{1+\theta}~\frac{1}{1-e^{-1}}~\sin^{-1}{\left(\frac{\pi}{d+1}\right)}. \end{eqnarray*}
Starting at equilibrium  $\eta_0=\pi$, for any function $f$ on $S$ we also have the uniform estimate
\begin{eqnarray*}
\sup_{n\geq 0}w_n^{{\tiny soft}}(f)&\leq& \pi\left(\left[f-\pi(f)\right]^2\right)
\displaystyle+~2~E_0~\frac{1+E_0}{1+\overline{E}_1}~u\left([f-\pi(f)]^2\right).
\end{eqnarray*}
\end{prop}
The proofs of (\ref{v-soft-varphi-ref}) and proposition~\ref{estimates-var-soft}  are provided in section~\ref{soft-obstacle-samplers}, on page~\pageref{proof-v-soft-varphi-ref}.

 \subsection{Particle sampler with hard obstacles}\label{psh-sec}

This particle scheme is based on sampling $N$ copies $\widehat{\xi}_n:=(\widehat{\xi}^i_n)_{1\leq N}$ 
of the absorbed random walk $X^c_n$. The transition of the chain is decomposed as above. When a particle is absorbed, instantly one of the remaining particles duplicates. We let $(\widehat{\xi}^i_n)_{1\leq N}$  be the particle system before the killing transition.
The corresponding Monte Carlo estimates are given by

\begin{eqnarray*}
\Za^{{\tiny hard}}_{n}&:=&\prod_{1\leq p\leq n}\widehat{\eta}^{{\tiny hard}}_{p}(1_S)\quad\mbox{\rm with}\quad
\widehat{\eta}^{{\tiny hard}}_{n}:=\frac{1}{N}
\sum_{1\leq i\leq N}~\delta_{\widehat{\xi}^i_n}\\
\eta^{{\tiny hard}}_{n}&:=&\Psi_{1_S}(\widehat{\eta}^{{\tiny hard}}_{n})\quad\mbox{\rm and}\quad
\gamma^{{\tiny hard}}_{n}:=
\Za^{{\tiny hard}}_{n}~\times~\eta^{{\tiny hard}}_{n}.
\end{eqnarray*}

The Monte Carlo approximation $\QQ^{{\tiny hard}}_{n}$ of $\QQ_{n}$ is defined as above by replacing ${\widehat{\xi}^i_n}$ by their ancestral lines. 

We have implicitly assumed that 
the state space of the particle sampler is the extended space $S\cup\{c\}$ and functions are null on the coffin state $c$. In this situation,
we use the convention $\eta^{{\tiny hard}}_{n}=\widehat{\eta}^{{\tiny hard}}_{n+1}=\delta_c$ as soon as all particles have exit the set $S$ at some earlier time horizon. 

The variance analysis of these particle estimates follows the same line of arguments as the ones of the soft-particle sampler. 
For a more thorough discussion on these variances we refer the reader to section~\ref{var-analysis-hard-bk}.

We also emphasize that the variances of these hard particle samplers are generally larger than the ones of the soft obstacles.
For instance we have
\begin{equation}\label{1st-comparision}
v_{n}^{{\tiny hard}}(f)\geq {v}_{n-1}^{{\tiny soft}}(Q(f))\quad
\mbox{\rm and}\quad  w_{n}^{{\tiny hard}}(f)\geq {w}^{{\tiny soft}}_{n-1}\left(\frac{Q}{\eta_{n-1}(Q(1))}
  \left[f-\eta_n(f)\right]\right).
\end{equation}
An explicit description of these variances and the proof of these inequalities are provided  in section~\ref{var-at-killing-sec}.

At equilibrium, for any function $f$ we have
\begin{equation}\label{equilibrium-hard-1}
w_{n}^{{\tiny hard}}(f)\geq E_0~w_{n}^{{\tiny hard}}(f)=E_0^{-1}~{v}^{{\tiny hard}}_{n}\left(
f-\pi(f)\right)=w_{n-1}^{{\tiny soft}}(f)=v^{{\tiny soft}}_{n-1}(f-\pi(f)).
\end{equation}
When $f=\varphi_0$ these variances have a simple expression 
\begin{equation}\label{equilibrium-hard-2}
\displaystyle E_0^{-1}~ v_{n}^{{\tiny hard}}(\varphi_0)
\displaystyle=\left[1+(n-1)~\left(1- E^2_0\right)\right]~\pi\left(\left[
\varphi_0-
\pi(\varphi_0)\right]^2\right)+n~\left[1-E_0\right]~\pi(\varphi_0)^2.
\end{equation}
The detailed proofs of  (\ref{equilibrium-hard-1})  and (\ref{equilibrium-hard-2}) are provided in section~\ref{var-equilibrium-sec}.

The unnormalized measure $\widehat{\gamma}_{n+1}$ discussed in (\ref{geometric-formula-hard}) can also be approximated as $N\uparrow\infty$  using the formula
$$
\widehat{\gamma}_{n+1}(f)= \gamma_n(K(f))~\simeq~\widehat{\gamma}^{{\tiny hard}}_{n+1}(f):=\Za^{{\tiny hard}}_{n}\times \widehat{\eta}^{{\tiny hard}}_{n+1}(f).
$$
The variance of these particle schemes are defined by
\begin{eqnarray}
\widehat{v}_{n+1}^{{\tiny hard}}(f)&=&\displaystyle\lim_{N\rightarrow\infty}N~\EE\left(\left[\Za^{{\tiny hard}}_{n}\Za^{-1}_{n}~\widehat{\eta}^{{\tiny hard}}_{n+1}(f)-\widehat{\eta}_{n+1}(f)\right]^2\right)\geq v_n^{{\tiny soft}}(K(f))\nonumber\\
\widehat{w}_{n+1}^{{\tiny hard}}(f)&=&\displaystyle\lim_{N\rightarrow\infty}N~\EE\left(\left[\widehat{\eta}^{{\tiny hard}}_{n+1}(f)-\widehat{\eta}_{n+1}(f)\right]^2\right)\geq w_{n}^{{\tiny soft}}(K(f)).\label{ref-comparison}
\end{eqnarray}

The detailed proofs of the formulae and the estimates stated above are provided in section~\ref{var-analysis-hard}.

\section{A brief review on Feynman-Kac semigroups}\label{sec:fk}
\subsection{Evolution equations}\label{review-FK-sec}
 
Let $\Phi^{n}(\eta_{0})=\eta_{n}$, with $n\geq 0$, be the semigroup of the equation defined in (\ref{def-Phi}).
The $n$-th power $Q^{n}$ of the matrix $Q$ introduced in (\ref{def-Q-theta}) can be interpreted as the Feynman-Kac semigroup defined
for any function $f$ on $S$ and any $x\in S$ by
\begin{eqnarray}
Q^{n}(f)(x)&:=&\EE\left(f(Y_n)~Z_n(Y)~|~Y_0=x\right)\nonumber\\
&=&\EE\left(f(X_n)~\prod_{1\leq p\leq n}1_{S}(X_p)~|~X_0=x\right)~~\Longrightarrow~\Phi^{n}(\eta_{0})=\frac{\eta_{0}Q^{n}}{\eta_{0}Q^{n}(1)}.
\label{def-Qpn}
\end{eqnarray}
In this notation the Markov operator $P_n$ defined in (\ref{def-Pn}) is given by
\begin{eqnarray*}
P_n(f)(x)&:=&\Phi^{n}(\delta_x)(f)={Q^{n}(f)(x)}/{Q^{n}(1)(x)}.\end{eqnarray*}
Also observe that the semigroup $Q^n$ can be interpreted as the conditional distributions
$$
Q^n(f)(x)=\EE\left(f(X_n)~1_{T_X>n}~|~X_0=x\right)=\EE\left(f(Y^c_n)~1_{T_Y\geq n}~|~Y_0=x\right).
$$
By proposition 12.1.7 in~\cite{delmoral2} for any probability measures $(\mu_1,\mu_2)$ on $S$ we have the Lipschitz inequality
\begin{eqnarray}
\Vert \Phi^{n}(\mu_1)-\Phi^{n}(\mu_2)\Vert_{\tiny tv}&\leq& \frac{\Vert Q^n(1)\Vert}{\max{\left(\mu_1Q^n(1),\mu_2Q^n(1)\right)}}~\beta(P_n)~\Vert \mu_1-\mu_2\Vert_{\tiny tv}\nonumber\\
&&\nonumber\\
&\leq &\rho( Q^n(1))~\beta(P_n)~\Vert \mu_1-\mu_2\Vert_{\tiny tv}.\label{contraction-fk-sg}
\end{eqnarray}

Recalling that $g=K(1_S)$, we readily check that the conditional distributions $\widehat{\eta}_n$ introduced in (\ref{geometric-formula-hard}) satisfy the evolution equation
\begin{equation}\label{def-Phi-hat}
 \widehat{\eta}_{n+1}=\widehat{\Phi}( \widehat{\eta}_{n}):=\Psi_{\widehat{g}}(\widehat{\eta}_{n})\widehat{M}\quad
\mbox{\rm with}\quad (\widehat{g},\widehat{M}):=(1_S,K)
\end{equation}
with the convention $\widehat{\eta}_0:=\eta_{-1}K=\eta_0$ for $n=0$.
To check this claim we use the decompositions
$$
 \widehat{\eta}_{n+1}(f)=\eta_{n}K(f)=\frac{\eta_{n-1}K(1_S~K(f))}{\eta_{n-1} K(1_S)}=
\frac{\widehat{\eta}_n(1_S~K(f))}{\widehat{\eta}_n(1_S)}=\Psi_{1_S}(\widehat{\eta}_n)K(f).
$$
Observe that (\ref{def-Phi-hat}) is defined as in (\ref{def-Phi}) by replacing $(g,M)$ by 
$(\widehat{g},\widehat{M})$. In the reverse angle, we have
$$
\Psi_{1_S}(\widehat{\eta}_n)=\Psi_{1_S}(\eta_{n-1}K)=\Psi_{g}(\eta_{n-1})M=\eta_n.
$$
We let $\widehat{\Phi}^{n}(\widehat{\eta}_{0})=\widehat{\eta}_{n}$, with $n\geq 0$, be the semigroup of the equation defined above. We also have the product formula
$$
\Za_{n}=\prod_{0\leq p<n}\eta_{p}(g)=\prod_{1\leq p\leq n}\widehat{\eta}_{p}(1_S)\quad\mbox{\rm and}\quad
\widehat{\Phi}^n(\widehat{\eta}_0)=
\frac{\widehat{\eta}_0\widehat{Q}^n}{\widehat{\eta}_0\widehat{Q}^n(1)}.
$$

\subsection{Spectral decompositions}\label{spectral-decomposition-sec}

The $(d\times d)$-symmetric matrices $Q$ defined in (\ref{def-Q-theta}) can be rewritten as
$$
Q=\frac{1}{2+\theta}\left(
\begin{array}{ccccccc}
\theta&1&0&0&0&\ldots&0\\
1&\theta&1&0&0&\ldots&0\\
0&1&\theta&1&0&\ldots&0\\
\vdots&&&\ldots&\ldots&\ldots&0\\
\vdots&&&\ldots&\ldots&\ldots&0\\
0&0&0&\ldots&1&\theta&1\\
0&0&0&\ldots&0&1&\theta
\end{array}
\right)=\mbox{\rm diag}\left(g(1),\ldots,g(d)\right)~M
$$
where $\mbox{\rm diag}\left(g(1),\ldots,g(d)\right)$ stands for the diagonal matrix with entries $g(x)$.
The matrix $M$ is clearly reversible w.r.t. the
probability measure
$
\mu=\Psi_{g}(u)
$. In addition, we have
$$
\pi:=\Psi_{\varphi_{0}}(u)\Longrightarrow \Phi(\pi)=\pi.
$$
The eigenvalues of the tridiagonal Toeplitz matrices $Q$ are defined for any $1\leq i\leq d$ by
\begin{eqnarray*}
E_{i-1}
&=&\frac{\theta+2~\cos{\left(\frac{i}{d+1}~\pi\right)}}{\theta+2}
\end{eqnarray*}
and the right $\LL_2(u)$-normalized eigenvectors
$$
\displaystyle\varphi_{i-1}(x)=\sqrt{
\frac{2d}{d+1}}~
\sin{\left[\frac{i}{d+1}~x~\pi\right]}.
$$
In this notation, we have the spectral decompositions
\begin{eqnarray}\label{spectral-decomp-Qn}
Q^{n}(x,y)&=&\sum_{0\leq i<d}~E_{i}^n~\varphi_{i}(x)~\varphi_{i}(y)~u(y).
\end{eqnarray}

The Doob $\varphi_{0}$-process, corresponding to the ground state eigenfunction $\varphi_{0}$ defined above,
is a Markov chain $Y^{\varphi}_n$ on $S$, with initial distribution $\eta^{\varphi}_0=\Psi_{\varphi_0}(\eta_0)$, and the Markov transition
\begin{equation}\label{ref-Qn-varphi}
\begin{array}{l}
\displaystyle M_{\varphi}(x,y):=E_0^{-1}\times \varphi_{0}^{-1}(x)~Q(x,y)~\varphi_{0}(y)=\frac{M(x,y)\varphi_{0}(y)}{M(\varphi_{0})(x)}\\
\\
\Longrightarrow Q^n(x,y)=\displaystyle E_0^n~\varphi_{0}(x)~M_{\varphi}^n(x,y)~\varphi_{0}(y)^{-1}\Rightarrow
P_n(x,y)=\displaystyle \frac{M_{\varphi}^n(x,y)~\varphi_{0}(y)^{-1}}{M_{\varphi}^n(\varphi_{0}^{-1})(x)}.
\end{array}
\end{equation}
This yields the estimate
\begin{equation}\label{ref-Pn-proof}
\beta(P_n)\leq \rho(\varphi_{0})~\beta(M_{\varphi}^n).
\end{equation}
Also observe that $M_{\varphi}$ is reversible w.r.t. the probability measure
$$\pi_{\varphi}:=\Psi_{\varphi_{0}M(\varphi_{0})}(\mu)=\Psi_{\varphi^2_{0}/g}(\mu)=\Psi_{\varphi^2_{0}}(u).
$$
In addition, we have the $\LL_2(\pi_{\varphi})$-spectral decomposition
$$
M_{\varphi}^n(x,y)=\pi_{\varphi}(y)+
\sum_{1\leq i<d}~\overline{E}_i^n~\overline{\varphi}_{i}(x)~\overline{\varphi}_{i}(y)~\pi_{\varphi}(y).
$$
with the orthonormal basis functions
$
\overline{\varphi}_{i}:=\varphi_{i}/\varphi_{0}$
 and the eigenvalues
 $
\overline{E}_i:=E_i/E_0
$.
Considering the $\LL_2(\pi_{\varphi})$ spectral decomposition of the function
$$
f=\pi_{\varphi}(f)+\sum_{1\leq i<d}~\pi_{\varphi}\left(f~\overline{\varphi}_{i}\right)~\overline{\varphi}_{i}
\Rightarrow
\pi_{\varphi}\left(\left[f-\pi_{\varphi}(f)\right]^2\right)=\sum_{1\leq <d}~\pi_{\varphi}\left(f~\overline{\varphi}_{i}\right)^2
$$ 
and
\begin{eqnarray*}
\pi_{\varphi}\left[\left(M_{\varphi}^n(f)-\pi_{\varphi}(f)\right)^2\right]&=&\sum_{1\leq i<d}~\overline{E}_i^{2n}~\pi_{\varphi}\left(f~\overline{\varphi}_{i}\right)^2\leq  \overline{E}_1^{2n}~\pi_{\varphi}\left(\left[f-\pi_{\varphi}(f)\right]^2\right).
\end{eqnarray*}
This ends the proof of the first assertion in theorem~\ref{stab-theorem}.

We end this section with a brief discussion on the matrices
$\widetilde{Q}$ introduced in (\ref{tilde-Q-def}).
Since ${Q}$ is symmetric, the matrix $\widetilde{Q}$ is $\Psi_{1/g}(u)$-reversible. This implies that
$$
\sqrt{\Psi_{1/g}(u)(x)}~\widetilde{Q}(x,y)~\frac{1}{\sqrt{\Psi_{1/g}(u)(y)}}=
\sqrt{\Psi_{1/g}(u)(y)}~\widetilde{Q}(y,x)~\frac{1}{\sqrt{\Psi_{1/g}(u)(x)}}
$$
or equivalently
$$
R(x,y):=
\sqrt{g(x)}~{Q}(x,y)~\sqrt{g(y)}~=\sqrt{g(y)}~{Q}(y,x)~\sqrt{g(x)}.
$$
This shows that
\begin{eqnarray*}
\sqrt{g(x)}~{Q}(x,y)~\sqrt{g(y)}&=&\sum_{0\leq i<d}~\widetilde{E}_{i}(\theta)^n~\psi_{i}(x)~
\psi_{i}(y)~u(y)
\end{eqnarray*}
for a non increasing sequence of ${R}$-eigenvalues $\widetilde{E}_{0}>\widetilde{E}_{1}\geq \ldots\geq \widetilde{E}_{d}$ (with
$\widetilde{E}_{d}>-\widetilde{E}_{0}$ and $\widetilde{E}_{0}\in ]0,1[$~) and some orthonormal basis of $\LL_2(u)$
of right eigenvectors $$
\sqrt{g}~{Q}(\sqrt{g}~{\psi}_{i})=\widetilde{E}_{i}~{\psi}_{i}.$$
This readily implies the $\LL_2(\Psi_{1/g}(u))$-spectral decomposition
\begin{eqnarray*}
\widetilde{Q}(x,y)&=&
\sum_{0\leq i<d}~\widetilde{E}_{i}^n~\widetilde{\psi}_{i}(x)~\widetilde{\psi}_{i}(y)~
\Psi_{1/g}(u)(y)
\end{eqnarray*}
with the eigenvectors
$$
\widetilde{\psi}_{i}:=
\sqrt{g~u(1/g)}~\psi_{i}\Rightarrow \Psi_{1/g}(u)\left(\widetilde{\psi}_{i}\widetilde{\psi}_{j}\right)=u({\psi}_{i}{\psi}_{j})=1_{i=j}.
$$

\subsection{Some quantitative estimates}\label{some-quantitative-sec}
We have
$$
\mu(x)=\frac{g(x)}{d-(1+\theta/2)^{-1}}\geq\inf_{x\in S}\mu(x)=\frac{1}{d}~\frac{1+\theta}{2(1-1/d) +\theta}\geq
 \frac{1+\theta}{2+\theta}~\frac{1}{d}\geq \frac{1}{2d}.
$$
It is also readily checked that
$$
\pi_{\varphi}(x)=\frac{2}{d+1}~\sin^2{\frac{x\pi}{d+1}}\geq 
\displaystyle\inf_{x\in S}{\pi_{\varphi}(x)}:= s_2^{-2}(d)\quad\mbox{\rm and}\quad
\rho(g)=\frac{2+\theta}{1+\theta}.
$$

\begin{prop}\label{proposition-ratio-functions}
We have the following uniform estimates
\begin{eqnarray*}
\sup_{n\geq 0}\rho(Q^{n}(1))&\leq& \frac{2+\theta}{1+\theta}~\sin^{-1}{\left(\frac{\pi}{d+1}\right)}
\quad\mbox{ and}\quad
\rho(\varphi_0)=\left\{
\begin{array}{rcl}
\displaystyle\frac{1}{\sin{\frac{\pi}{d+1}}}&\mbox{for odd}& q\\
&&\\
\displaystyle\cot{\frac{\pi}{d+1}}&\mbox{for even}& q 
\end{array}
\right\}.\end{eqnarray*}
\end{prop}
The proof of the proposition is rather technical thus it is housed in the appendix.

The estimate (\ref{Qn1-rho-ref}) is a direct consequence of the uniform estimate stated in proposition~\ref{proposition-ratio-functions}.
By (\ref{ref-Qn-varphi}) we have the exponential decays of the function
$n\mapsto Q^n(1)(x)$ 
are given by
\begin{equation}\label{ref-E0Qn1}
\rho(\varphi_0)^{-1}~E_0^n \leq Q^n(1)\leq \rho(\varphi_0)~E_0^n.
\end{equation}
 The proof of (\ref{Qn1-ref}) is also a direct 
 consequence of (\ref{ref-E0Qn1}) and the estimate of $\beta\left(M_{\varphi}^n\right)$ 
 stated in  theorem~\ref{stab-theorem}.

By proposition 3 in~\cite{diaconis-stroock} we have 
 \begin{eqnarray*}
\Vert\delta_xM_{\varphi}^{n}-\pi_{\varphi}\Vert_{\tiny tv}&\leq&
 \frac{1}{2}~\overline{E}_1^{\,n}~\left(\pi_{\varphi}(x)\right)^{-1/2}\leq \frac{1}{2}~\overline{E}_1^{\,n}~\sqrt{\frac{d+1}{2}}~\frac{1}{\sin{\frac{\pi}{d+1}}}.
 \end{eqnarray*}

 Now we are in position  to prove theorem~\ref{stab-theorem} and theorem~\ref{theo-stab-Pn}.\\
 
 {\bf Proof of theorem~\ref{stab-theorem}:}
 
 The estimate stated above yields the estimate of the Dobrushin contraction coefficient $\beta\left(M_{\varphi}^n\right)$ stated in the r.h.s. of (\ref{L2+beta}). The l.h.s. of (\ref{contract+pi}) is now a consequence of (\ref{def-Pn-ineq})  and the estimate  (\ref{ref-Pn-proof}).
 The proof of the contraction inequality in the r.h.s. of (\ref{contract+pi})  is a simple combination of the l.h.s. of (\ref{contract+pi})  with
the contraction inequality (\ref{contraction-fk-sg}) and the estimate  (\ref{ref-Pn-proof}). This ends the proof of theorem~\ref{stab-theorem}\cqfd

 {\bf Proof of theorem~\ref{theo-stab-Pn}:}
 
  The estimate (\ref{betaPn}) is a consequence of (\ref{ref-Pn-proof}) and the estimates of $\rho(\varphi_0)$ and $\beta\left(M_{\varphi}^n\right)$ 
 stated in proposition~\ref{proposition-ratio-functions} and in theorem~\ref{stab-theorem}. 
  
 Now we come to the proof of (\ref{def-varsigmaP}). 
 Firstly, we use the decomposition
 $$
  \delta_xP_{m+n}=\Psi_{M^m_{\varphi}(\varphi_0^{-1})}\left(\delta_xM^n_{\varphi}\right)P_m=
  \Psi_{Q^m(1)/\varphi_0}\left(\delta_xM^n_{\varphi}\right)P_m
 $$
to check that
 $$
 \delta_xP_{m+n}-\delta_yP_{m+n}=\left[ \Psi_{Q^m(1)/\varphi_0}\left(\delta_xM^n_{\varphi}\right)-\Psi_{Q^m(1)/\varphi_0}\left(\delta_yM^n_{\varphi}\right)\right]P_m.
 $$
This implies that
\begin{eqnarray*}
\beta(P_{m+n})&\leq& \left[\rho_Q~\rho(\varphi_0)~\beta(M^n_{\varphi})\right]~\beta(P_{m})\leq  \left[
\frac{2+\theta}{1+\theta}~s_3(d)~\overline{E}_1^{\,n}\right]~\beta(P_{m}).
\end{eqnarray*}
The end of the proof of the l.h.s. of  (\ref{def-varsigmaP}) is now clear.
This also implies that
\begin{eqnarray*}
\sum_{n\geq 0}\beta(P_n)^2&=&\sum_{k\geq 0}~\sum_{k\,\varsigma_{P}\leq n<(k+1)\,\varsigma_{P}}\beta(P_n)^2\leq \varsigma_{P}~\sum_{k\geq 0}e^{-2k}\leq \frac{1}{1-e^{-2}}~\varsigma_{P}.
\end{eqnarray*}
This ends the proof of (\ref{def-varsigmaP}). 
Also observe that
\begin{eqnarray*}
\frac{1}{\log{\left(1+\frac{1-\overline{E}_1}{\overline{E}_1}\right)}}
&\leq& \frac{\overline{E}_1}{1-\overline{E}_1}~\frac{1}{1-\frac{1}{2}\left(\frac{1-\overline{E}_1}{\overline{E}_1}\right)}\leq \sin^{-2}{\left(\frac{1}{2}~\frac{\pi}{d+1}\right)}~(1+\theta/2)
\end{eqnarray*}
as soon as $d$ is chosen so that
$$
\frac{1}{2}~\frac{1-\overline{E}_1}{\overline{E}_1}=\frac{2~\sin{\left(\frac{3}{2}~\frac{\pi}{d+1}\right)}\sin{\left(\frac{1}{2}~\frac{\pi}{d+1}\right)}}{\theta+2\cos{\frac{2\pi}{d+1}}}~< 1/2.
$$
This condition is met as soon as
$$
\frac{3}{2}\left(\frac{\pi}{d+1}\right)^2<\theta/2+\left(1-2\left(\frac{\pi}{d+1}\right)^2\right)
\left(\leq \theta/2+\left(1-2\sin^2{\frac{\pi}{d+1}}\right)\right).
$$
This yields the sufficient condition
$$
d\geq 2\pi-1>\pi~\sqrt{\frac{7}{2}}-1\left(\geq \pi~\sqrt{\frac{7}{2}~\frac{1}{1+ \theta/2}}-1\right).
$$
This ends the proof of theorem~\ref{theo-stab-Pn}.\cqfd

We end this section with the spectral analysis of the matrices
$$
{R}=\frac{1}{2+\theta}\left(
\begin{array}{ccccccc}
\theta ~\epsilon_{\theta}&\sqrt{\epsilon_{\theta}}&0&0&0&\ldots&0\\
\sqrt{\epsilon_{\theta}}&\theta&1&0&0&\ldots&0\\
0&1&\theta&1&0&\ldots&0\\
\vdots&&&\ldots&\ldots&\ldots&0\\
\vdots&&&\ldots&\ldots&\ldots&0\\
0&0&0&\ldots&1&\theta&\sqrt{\epsilon_{\theta}}\\
0&0&0&\ldots&0&\sqrt{\epsilon_{\theta}}&\theta~\epsilon_{\theta}
\end{array}
\right)
\quad\mbox{\rm with}
\quad
\frac{1}{2}\leq 
\epsilon_{\theta}=\frac{1+\theta}{2+\theta}\leq 1.
$$
By the Perron-Frobenius theorem we have $\rho({\psi}_{0})\vee \rho(\widetilde{\psi}_{0})<\infty$. In addition, arguing as in the proof of (\ref{ref-Qn-varphi})
and (\ref{ref-E0Qn1}) we have
\begin{equation}\label{tilde-E-Q1}
\rho(\widetilde{\psi}_{0})^{-1}~\widetilde{E}_0^n \leq \widetilde{Q}^n(1)\leq \rho(\widetilde{\psi}_{0})~\widetilde{E}_0^n.
\end{equation}
The spectral inequalities stated in theorem~\ref{spectral-comparison-theo} readily implies that
$$
 (\rho(\widetilde{\psi}_{0})\rho(\varphi_0)^2)^{-1}~({\widetilde{E}_0}/{E_0^{2}})^n\leq 
\frac{\eta_0\widetilde{Q}^n(1)}{[\eta_0{Q}^n(1)]^2}\leq \rho(\widetilde{\psi}_{0})\rho(\varphi_0)^2~({\widetilde{E}_0}/{E_0^{2}})^n.$$
This ends the proof of (\ref{var-tilde-inequality}).

\subsection{Some combinatorial properties}
Let  $Q_0$ be the matrix associated with 
the null parameter $\theta=0$. 
The computation of the entries of the power matrices $Q_0^n$ resumes to the computation of the number
of paths of the simple random walk confined in the set $S$. More precisely, we have
$$
Q^{n}_0(x_0,x_n)=2^{-n}~C_n(x_0,x_n)
$$
where $C_n(x_0,x_n)$ stands for the number of paths of length $n$ in $S$ of the simple random walk 
starting at $x_0$ and ending at $x_n$; that is, we have that
$$
C_n(x_0,x_n):=\mbox{\rm Card}\left\{(x_1,\ldots,x_{n-1})\in S^{n-1}~:~\forall 1\leq k\leq n~~\vert x_k-x_{k-1}\vert=1~\right\}.
$$
The powers of the matrices $Q$ and $Q_0$ are related by the formula 
\begin{eqnarray}\label{spectral-decomp-Qn-counting}
Q^{n}(x,y)&=&\frac{1}{(1+\theta/2)^n}~\sum_{0\leq m\leq n}\left(
\begin{array}{c}
n\\m
\end{array}
\right)~\left(\theta/2\right)^m~Q^m_0(x,y).
\end{eqnarray}

We let $C_n(x)$ be the number of non absorbed paths starting from $x$; that is
$$
C_n(x):=\sum_{y\in S}C_n(x,y).
$$
By (\ref{spectral-decomp-Qn}) we have the formulae
\begin{eqnarray*}
Q^{n}_0(x,y)&=&2^{-n}~C_n(x,y)\\
&=& \frac{2}{d+1}\sum_{1\leq i\leq d}~\cos^n{\left(\frac{i}{d+1}~\pi\right)}~\sin{\left[\frac{i}{d+1}~x~\pi\right]}~\sin{\left[\frac{i}{d+1}~y~\pi\right]}
\end{eqnarray*}
and
\begin{eqnarray*}
Q^{n}_0(1)(x)&=&2^{-n}~C_n(x)\\
&=& \frac{2}{d+1}\sum_{1\leq i~\mbox{\tiny odd}~\leq d}~\cos^n{\left(\frac{i}{d+1}~\pi\right)}~\sin{\left[\frac{i}{d+1}~x~\pi\right]}~\cot{\left[\frac{i}{d+1}~\pi\right]}.
\end{eqnarray*}
Using the sine multiple-angle formula
$$
\sin{(x\alpha)}=\sin{(\alpha)}~\sum_{0\leq l\leq \lfloor (x-1)/2\rfloor}~(-1)^l~\left(
\begin{array}{c}
x-(l+1)\\
l
\end{array}
\right)~2^{x-2l-1}~\cos^{x-2l-1}{\left(\theta\right)}
$$
which is valid for any $x\in\NN$ and $\alpha\in\RR$, we find that
\begin{eqnarray*}
C_n(x)
&=& \sum_{0\leq l\leq \lfloor (x-1)/2\rfloor}~(-1)^l~\left(
\begin{array}{c}
(x-1)-l\\
l
\end{array}
\right) ~ C_{n+(x-1)-2l}(1).
\end{eqnarray*}
This gives a simple proof of the recurrence formula presented in~\cite{everett,krewaras} using sophisticated combinatorial and spectral techniques.

\section{Variance analysis}\label{var-analysis}

This section provides a detailed discussion on the variance of the Sequential Monte Carlo samplers presented in section~\ref{smc-section}.
We shall not discuss the variance of the importance sampling technique based on reflected processes discussed in section~\ref{twisted-reflected-sec}. The degeneracy of these Monte Carlo schemes w.r.t. the time horizon has been already dicussed in  theorem~\ref{spectral-comparison-theo}.

Most of the section is dedicated with the proofs of the variance estimates of the hard and soft particle samplers 
stated in section~\ref{smc-section}. These variances are expressed in terms
of the Feynman-Kac linear and non linear semigroups $(Q^n,\eta_n)$ and $(\widehat{Q}^n,\widehat{\eta}_n)$ discussed in section~\ref{description-models-sec} and section~\ref{review-FK-sec}. Explicit formulae can be derived when the system start at equilibrium, that is when $\eta_0=\pi$. We shall also discuss the situation where the test function $f$ is given by the eigenfunction $\varphi_0$. 

In connection to this, we already notice that
$$
\pi=\Psi_{\varphi_0}(u)\Rightarrow
\pi(\varphi_0)={1}/{u(\varphi_0)}=\pi(1/\varphi_0)=\sqrt{\frac{d(d+1)}{2}}~\tan{\left(\frac{1}{d+1}~\frac{\pi}{2}\right)}.
$$
This elementary observation can be used to derive explicit variance formulae for the importance sampling schemes based on the twisted $\varphi_0$-Doob process discussed in section~\ref{twisted-section-dp}. 
For instance using (\ref{var-formula-dp}) it is readily check that
$$
v_n^{{\tiny dp}}(\varphi_0)=\pi(\varphi_0)~(1-\pi(\varphi_0))=\frac{1}{u(\varphi_0)}\left(\frac{1}{
u(\varphi_0)}-1\right)
$$
and
$$
w_n^{{\tiny dp}}(\varphi_0)=\pi(\varphi_0)^2~\left(\pi(\varphi_0)\pi(1/\varphi_0)-1\right)=\frac{1}{u(\varphi_0)^2}\left(\frac{1}{
u(\varphi_0)^2}-1\right).
$$

At equilibrium the normalized Feynman-Kac semigroups (\ref{normalized-fksg}) become
$$
\eta_0=\pi\Longrightarrow{Q}_{p,p+n}(x,y)=\frac{Q^n(x,y)}{\eta_0(Q^n(1))}=E_0^{-n}Q^n(x,y)=E_0^{-n}Q^n(1)(x)~P_n(x,y).
$$
In addition, by (\ref{Qn1-rho-ref}) we have the uniform estimate
$$
E_0^{-n}Q^n(1)(x)=
\frac{Q^n(1)(x)}{\pi(Q^n(1))}\leq \rho(Q^n(1))\leq \frac{2+\theta}{1+\theta}~\sin^{-1}{\left(\frac{\pi}{d+1}\right)}.
$$

\subsection{The soft obstacle model}\label{soft-obstacle-samplers}
This section is dedicated to the analysis of the variances of the particle sampler with reflection presented in section~\ref{psr-sec}.
By (\ref{v-soft-f}) we  have the variance formula
\begin{eqnarray}
\displaystyle w_n^{{\tiny soft}}(f)
&=&v_n^{{\tiny soft}}(f-\eta_n(f))\nonumber\\
&=&\eta_n\left(\left[f-\eta_n(f)\right]^2\right)
\displaystyle+
\sum_{0\leq p<  n}~(1-\eta_{p}(g)^2)~
\eta_{p}\left(\left[{Q}_{p,n}(f-\eta_n(f))\right]^2\right)\label{ref2-hs}.
\end{eqnarray}
At equilibrium $\eta_0=\pi$ these formulae become
\begin{eqnarray*}
\displaystyle w_n^{{\tiny soft}}(f)&=&\pi\left(\left[f-\pi(f)\right]^2\right)
\displaystyle+~(1-E_0^2)~
\sum_{1\leq p\leq   n}
\pi\left(\left[\frac{{Q}^{p}}{E_0^{p}}(f-\pi(f))\right]^2\right)\\
\displaystyle v_n^{{\tiny soft}}(f)&=&\sum_{0\leq p\leq   n}\pi\left(\left[\frac{{Q}^{p}(f)}{E_0^{p}}-\pi(f)\right]^2\right)-~E_0^2~\sum_{1\leq p\leq  n}
\pi\left(\left[~\frac{{Q}^{p}(f)}{E_0^{p}}-\frac{g}{E_0}~\pi(f)\right]^2\right) \\
&=&\sum_{0\leq p\leq   n}\pi\left(\left[\frac{{Q}^{p}(f)}{E_0^{p}}-\pi(f)\right]^2\right)-~\sum_{1\leq p\leq  n}
\pi\left(\left[~Q\left\{\frac{{Q}^{p-1}(f)}{E_0^{p-1}}-\pi(f)\right\}\right]^2\right).
\end{eqnarray*}
We have
$$
 v_n^{{\tiny soft}}(f)\leq \sum_{0\leq p\leq   n}\pi\left(\left[\frac{{Q}^{p}(f)}{E_0^{p}}-\pi(f)\right]^2\right).
$$
Also observe that
\begin{eqnarray*}
\displaystyle v_n^{{\tiny soft}}(f)
&\geq &\sum_{0\leq p\leq   n}\pi\left(\left[\frac{{Q}^{p}(f)}{E_0^{p}}-\pi(f)\right]^2\right)-E_0~\sum_{0\leq p<  n}
\pi\left(\left[\frac{{Q}^{p}(f)}{E_0^{p}}-\pi(f)\right]^2\right)\\
&=&\pi\left(\left[\frac{{Q}^{n}(f)}{E_0^{n}}-\pi(f)\right]^2\right)+(1-E_0)~\sum_{0\leq p<  n}
\pi\left(\left[\frac{{Q}^{p}(f)}{E_0^{p}}-\pi(f)\right]^2\right)
\end{eqnarray*}
This assertion follows from the fact that
\begin{eqnarray*}
\pi\left(\left[Q(f)\right]^2\right)&=&\pi\left(\left[K(1_Sf)\right]^2\right)\leq
\pi Q\left(f^2\right)=\pi(g)~\pi\left(f^2\right)=E_0~\pi\left(f^2\right).
\end{eqnarray*}
For instance when $\eta_0=\pi$ and $f=\varphi_0$ we have (\ref{v-soft-varphi-ref}).\label{proof-v-soft-varphi-ref}
Now we come to the proof of the variance estimated stated in proposition~\ref{estimates-var-soft}.\\

{\bf Proof of proposition~\ref{estimates-var-soft}:}

Starting at equilibrium we also have that
\begin{eqnarray*}
\pi\left(\left[\frac{{Q}^{n-p}}{E_0^{n-p}}(f-\pi(f))\right]^2\right)&=&\pi\left(\varphi_0^2~\left(M^{n-p}_{\varphi}\left([f-\pi(f)]/\varphi_0\right)\right)^2\right)\\
&=&\pi(\varphi_0)~\pi_{\varphi}\left(\varphi_0~\left(M^{n-p}_{\varphi}\left([f-\pi(f)]/\varphi_0\right)\right)^2\right).
\end{eqnarray*}
Recalling that
$$
\pi_{\varphi}\left([f-\pi(f)]/\varphi_0\right)=\frac{1}{\pi(\varphi_0)}~\pi\left([f-\pi(f)]\right)=0
$$
and $\Vert \varphi_0\Vert\leq \sqrt{2}$, for any $\mbox{\footnotesize osc}(f)\leq 1$ we find that
\begin{eqnarray*}
\pi\left(\left[\frac{{Q}^{n-p}}{E_0^{n-p}}(f-\pi(f))\right]^2\right)
&\leq &2~\overline{E}_1^{\,2(n-p)}~\pi_{\varphi}\left([f-\pi(f)]^2/\varphi_0^2\right)\\
&=&2~\overline{E}_1^{\,2(n-p)}~\frac{1}{u(\varphi_0^2)}
u\left([f-\pi(f)]^2\right)~\left(\leq 2~\overline{E}_1^{\,2(n-p)}\right)\\
&=&2~\overline{E}_1^{\,2(n-p)}~\left[
u\left([f-u(f)]^2\right)+\left(\pi(f)-u(f)\right)^2\right].
\end{eqnarray*}
This implies that
\begin{eqnarray*}
w_n^{{\tiny soft}}(f)
&\leq& \pi\left(\left[f-\pi(f)\right]^2\right)
\displaystyle+~2~\frac{1+E_0}{1+\overline{E}_1}~\frac{1-E_0}{1-\overline{E}_1}~u\left([f-\pi(f)]^2\right)\\
&\leq& \pi\left(\left[f-\pi(f)\right]^2\right)
\displaystyle+~2E_0~\frac{1+E_0}{1+\overline{E}_1}~u\left([f-\pi(f)]^2\right)\\
&\leq& \pi\left(\left[f-\pi(f)\right]^2\right)
\displaystyle+~4~u\left([f-\pi(f)]^2\right).
\end{eqnarray*}
The last assertion follows from
$$
\frac{1-E_0}{1-\overline{E}_1}=E_0~\frac{\sin{\left(\frac{\pi}{2(d+1)}\right)}}{\sin{\left(\frac{3}{2}~\frac{\pi}{d+1}\right)}}\leq E_0.
$$
More generally, observe that
$$
  \begin{array}{l}
\displaystyle
\left[\frac{{Q}^{n-p}}{\eta_{p}{Q}^{n-p}(1)}(f-\eta_n(f))\right](x)\\
\\
\displaystyle=\frac{{Q}^{n-p}(1)(x)}{\eta_{p}{Q}^{n-p}(1)}~\sum_{x^{\prime}\in S}~\eta_p(x^{\prime})~\frac{{Q}^{n-p}(1)(x^{\prime})}{\eta_{p}{Q}^{n-p}(1)}
\left[P_{n-p}(f)(x)-P_{n-p}(f)(x^{\prime})\right].
  \end{array}
  $$
  For any function $f$ with at most unit oscillations we find that
$$
\displaystyle
\left[\frac{{Q}^{n-p}}{\eta_{p}{Q}^{n-p}(1)}(f-\eta_n(f))\right]^2(x)
\displaystyle\leq \frac{{Q}^{n-p}(1)(x)}{\eta_{p}{Q}^{n-p}(1)}~\rho({Q}^{n-p}(1))~\beta(P_{n-p})^2.
  $$
  This shows that
  \begin{eqnarray*}
\displaystyle
\left\Vert\frac{{Q}^{n-p}}{\eta_{p}{Q}^{n-p}(1)}(f-\eta_n(f))\right\Vert&\leq& \rho({Q}^{n-p}(1))~\beta(P_{n-p})^2\\
&\leq&\frac{2+\theta}{1+\theta}~\sin^{-1}{\frac{\pi}{d+1}}~\beta(P_{n-p})^2.
  \end{eqnarray*}
Using (\ref{def-varsigmaP}) we conclude that
 \begin{eqnarray*}
 w_n^{{\tiny soft}}(f)&\leq& 1+
\displaystyle\frac{2}{1+\theta}~\sin^{-1}{\left(\frac{\pi}{d+1}\right)}
\sum_{0\leq p<  n}~
~\beta(P_{n-p})^2\\
&\leq &\displaystyle1+\frac{2}{1+\theta}~\frac{1}{1-e^{-1}}~\sin^{-1}{\left(\frac{\pi}{d+1}\right)}~(\varsigma_{P}-1).
 \end{eqnarray*}
 For path space models we also have
 $$
  w_n^{{\tiny\bf soft}}({f_n})\leq 1+\frac{2n}{1+\theta}~\frac{1}{1-e^{-1}}~\sin^{-1}{\left(\frac{\pi}{d+1}\right)}
 $$
  $$
v_n^{{\tiny soft}}(f)\leq (n+1)~\frac{2+\theta}{1+\theta}~\left[\sin{\left(\frac{\pi}{d+1}\right)}\right]^{-1}~\Vert f\Vert.
  $$
  This ends the proof of the proposition.\cqfd

\subsection{The hard obstacle model}\label{var-analysis-hard}

This section is dedicated to the analysis of the variances of the particle sampler with hard obstacles presented in section~\ref{psh-sec}.
Observe that the particle sampler $\widehat{\xi}_n$ is defined as $\xi_{n}$ by replacing $(g,M)$ by $(1_S,K)$. This simple observation
shows that the variance formulae in the hard obstacle case can be deduced from the ones of the soft obstacle sampler by replacing
 the Feynman-Kac semigroups associated to $(g,M)$ by the ones defined in terms of $(1_S,K)$. These variances are discussed in section~\ref{var-analysis-hard-bk}. We also show that these variances are larger than the ones associated with the soft obstacle sampler.

  \subsubsection{Variances before killing}\label{var-analysis-hard-bk}
  
As in (\ref{normalized-fksg}) and (\ref{v-soft-f}) the variance of these estimates are expressed in terms of the normalized semigroup
\begin{equation}\label{normalized-fksg-hard}
\widehat{Q}_{p,n}:=\frac{\widehat{Q}^{n-p}}{\widehat{\eta}_{p}\widehat{Q}^{n-p}(1)}~\Longrightarrow~\widehat{\eta}_p\widehat{Q}_{p,n}=\widehat{\eta}_n.
\end{equation}
For $n=p+1$ we have
$$
\widehat{Q}_{p,p+1}(f)=\frac{\widehat{Q}(f)}{\widehat{\eta}_{p}(1_S)}=\frac{1_S}{\widehat{\eta}_{p}(1_S)}~K(f).
$$

Replacing in (\ref{v-soft-f}) and (\ref{ref2-hs}) the quantities $(\eta_n,Q^n)$ by $
( \widehat{\eta}_n, \widehat{Q}^n)
  $ we have
  \begin{eqnarray*}
\displaystyle\widehat{v}_{n}^{{\tiny hard}}(f)
\displaystyle&=&\sum_{0\leq p\leq  n} \widehat{\eta}_p\left(\left[\widehat{Q}_{p,n}(f)- 
\widehat{\eta}_{n}(f)\right]^2\right)-
\displaystyle\sum_{0\leq p<n}~ 
 \widehat{\eta}_{p}\left(1_S~\left[K\widehat{Q}_{p+1,n}(f)-\widehat{\eta}_{n}(f)\right]^2\right)
 \end{eqnarray*}
as well as
  \begin{eqnarray*}
\displaystyle \widehat{w}_{n}^{{\tiny hard}}(f)
&=&\widehat{v}_{n}^{{\tiny hard}}(f-\widehat{\eta}_{n}(f))\\
&=&\widehat{\eta}_{n}\left(\left[f-\widehat{\eta}_{n}(f)\right]^2\right)
\displaystyle+
\sum_{1\leq p< n}~(1-\widehat{\eta}_{p}(1_S)^2)~
\widehat{\eta}_{p}\left(\left[\widehat{Q}_{p,n}(f-\widehat{\eta}_{n}(f))\right]^2\right) .
 \end{eqnarray*}
 
 We are now in position to prove the inequalities (\ref{ref-comparison}).
 
 {\bf Proof of formulae (\ref{ref-comparison}):}

  We have the commutation property
  $$
  \widehat{Q}^n(f)=1_S~Q^{n-1}(K(f))\Longleftrightarrow 1_S~K~  \widehat{Q}^n(f)= \widehat{Q}^{n+1}(f)=1_S~Q^n(K(f)).
  $$
  This implies
  that
  $$
    1_SK\widehat{Q}_{1,n}(f)=1_S~\frac{K\widehat{Q}^{n-1}(f)}{\eta_0 K \widehat{Q}^{n-1}(1)}=\frac{\widehat{Q}^{n}(f)}{\eta_0 \widehat{Q}^{n}(1)}=  \widehat{Q}_{0,n}(f).
  $$
Recalling that $ \widehat{\eta}_{0}= {\eta}_{0}$ has a support on $S$ we conclude that
  \begin{eqnarray*}
\widehat{\eta}_0\left(\left[\widehat{Q}_{0,n}(f)- 
\widehat{\eta}_{n}(f)\right]^2\right)&=&
 \widehat{\eta}_{0}\left(1_S~\left[K\widehat{Q}_{1,n}(f)-\widehat{\eta}_{n}(f)\right]^2\right).
 \end{eqnarray*}
 When $1\leq p<n$ we also have
 $$
 \widehat{Q}_{p,n}(f)=\frac{ \widehat{Q}^{n-p}(f)}{\eta_{p-1}K\widehat{Q}^{n-p}(1)}=
 1_S~\frac{{Q}^{n-p-1}(K(f))}{\eta_{p-1}{Q}^{n-p}(1)}
 $$
 and
 $$
1_S K \widehat{Q}_{p+1,n}(f)=\frac{\widehat{Q}^{n-p}(f)}{\eta_{p}K\widehat{Q}^{n-p-1}(1)}=1_S~
\frac{{Q}^{n-p-1}(K(f))}{\eta_{p}{Q}^{n-p-1}(1)}=1_S~Q_{p,n-1}(K(f)).
 $$
 Recalling that the measures $ {\eta}_{p}$ have a support on $S$ we prove that
 $$
  \begin{array}{l}
\displaystyle\widehat{v}_{n}^{{\tiny hard}}(f)\\
\\
\displaystyle= {\eta}_{n-1}K\left(\left[f- 
{\eta}_{n-1}(K(f))\right]^2\right)\\
\\
\hskip2cm\displaystyle+\sum_{0\leq p<  (n-1)} {\eta}_{p}K\left(\left[1_S~\frac{{Q}^{(n-1)-(p+1)}(K(f))}{
\eta_{p}{Q}^{(n-1)-p}(1)}- 
{\eta}_{n-1}(K(f))\right]^2\right)\\
\\
\hskip3cm\displaystyle-\displaystyle\sum_{1\leq p<n}~
 {\eta}_{p-1}Q\left(\left[
Q_{p,n-1}(K(f))- 
{\eta}_{n-1}(K(f))\right]^2\right).
  \end{array}
  $$
  Observe that for any measure $\mu$ and any function $f$ on $S$ we have
  $$
\mu(\left[Q(f)\right]^2)=\mu(\left[K(1_Sf)\right]^2) \leq \mu(K(1_Sf^2))= \mu Q(f^2).
  $$
  Applying Cauchy-Schwartz inequality, we conclude that
   $$
  \begin{array}{l}
\displaystyle\widehat{v}_{n}^{{\tiny hard}}(f)
\displaystyle\geq \sum_{0\leq p\leq   (n-1)} {\eta}_{p}\left(\left[ Q_{p,n-1}(K(f))- 
{\eta}_{n-1}(K(f))\right]^2\right)\\
\\
\hskip2cm\displaystyle-\displaystyle\sum_{1\leq p<n}~
 {\eta}_{p-1}\left(\left[~Q\left\{
Q_{p,n-1}(K(f))- 
{\eta}_{n-1}(K(f))\right\}\right]^2\right)={v}_{n-1}^{{\tiny soft}}(K(f)).
  \end{array}
  $$
  This also shows that
  $$
 \widehat{w}_{n}^{{\tiny hard}}(f)
=\widehat{v}_{n}^{{\tiny hard}}(f-\widehat{\eta}_{n}(f))\geq {v}_{n-1}^{{\tiny soft}}\left(\left[K(f)-{\eta}_{n-1}(K(f))\right]
\right)={w}_{n-1}^{{\tiny soft}}(K(f)).
  $$
    This ends the proof of the estimates (\ref{ref-comparison}).
  \cqfd
  
   \subsubsection{Variances at killing times}\label{var-at-killing-sec}
 After the killing transition the particle measures are given by
 $$
  \gamma^{{\tiny hard}}_{n}(f):=
\Za^{{\tiny hard}}_{n-1}~\times~\widehat{\eta}^{{\tiny hard}}_{n}(1_Sf)=\widehat{\gamma}^{{\tiny hard}}_{n}(f1_S)
 \quad\mbox{\rm 
 and}\quad
  \eta^{{\tiny hard}}_{n}(f)=\Psi_{1_S}\left(\widehat{\eta}^{{\tiny hard}}_{n}\right).
 $$
Therefore the variances of these empirical measures can be expressed in terms of the ones of the particle
sampler before killing. More precisely, we have the formulae
 $$
v_{n}^{{\tiny hard}}(f)=\widehat{v}_{n}^{{\tiny hard}}(f1_S) \quad\mbox{\rm 
 and}\quad   w^{{\tiny hard}}_{n}(f)=\widehat{w}^{{\tiny hard}}_{n}\left(\frac{1_S}{\widehat{\eta}_{n}(1_S)}
  \left[f-\Psi_{1_S}\left(\widehat{\eta}_{n}\right)(f)\right]\right).
 $$
 Observe that
 $$
 w^{{\tiny hard}}_{n}(f)={v}^{{\tiny hard}}_{n}\left(\frac{1}{{\eta}_{n-1}(g)}
  \left[f-{\eta}_{n}(f)\right]\right).
 $$
The formulae (\ref{1st-comparision}) are now a direct consequence of  (\ref{ref-comparison}).

We also have
 $$
  \begin{array}{l}
\displaystyle v_{n}^{{\tiny hard}}(f)
\displaystyle=\sum_{0\leq p<  n} {\eta}_{p}K\left(\left[1_S~\frac{{Q}^{(n-1)-p}(f)}{
\eta_{p}{Q}^{(n-1)-p}(1)}- 
{\eta}_{n-1}(Q(f))\right]^2\right)\\
\\
\hskip3cm\displaystyle-\displaystyle\sum_{1\leq p<n}~\eta_{p-1}(g)~
 {\eta}_p\left(\left[
Q_{p,n-1}(Q(f))- 
{\eta}_{n-1}(Q(f))\right]^2\right)
  \end{array}
  $$
and
$$
  \begin{array}{l}
w^{{\tiny hard}}_{n}(f)
\\
\\
=\displaystyle\widehat{w}^{{\tiny hard}}_{n}\left(\frac{1_S}{\widehat{\eta}_{n}(1_S)}
  \left[f-\Psi_{1_S}\left(\widehat{\eta}_{n}\right)(f)\right]\right)\\
\\
=\displaystyle\widehat{\eta}_{n}\left(1_S~\left[\frac{1}{{\eta}_{n-1}(g)}
  \left[f-{\eta}_{n}(f)\right]\right]^2\right)
\displaystyle+
\sum_{1\leq p< n}~(1-{\eta}_{p-1}(g)^2)~
\widehat{\eta}_{p}\left(\left[\widehat{Q}_{p,n}\left\{\frac{1_S}{\widehat{\eta}_{n}(1_S)}
  \left[f-{\eta}_{n}(f)\right]\right\}\right]^2\right).
  \end{array}
  $$
  Recalling that
  $$
  \widehat{Q}_{p,n}\left[\frac{1_S~f}{\widehat{\eta}_{n}(1_S)}\right]
  =\frac{1}{{\eta}_{n-1}(g)}~\frac{1_S~{Q}^{n-p}(f)}{{\eta}_{p-1}{Q}^{n-p}(1)
  }
  $$
  we conclude that
  $$
  \begin{array}{l}
{\eta}_{n-1}(g)~w^{{\tiny hard}}_{n}(f)
\\
\\
={\eta}_{n}\left(
  \left[f-{\eta}_{n}(f)\right]^2\right)
\displaystyle+\frac{1}{{\eta}_{n-1}(g)}~
\sum_{1\leq p< n}~(1-{\eta}_{p-1}(g)^2)~
\widehat{\eta}_{p}\left(1_S\left[
\frac{{Q}^{n-p}(f-\eta_n(f))}{{\eta}_{p-1}{Q}^{n-p}(1)
  }\right]^2\right) \\
  \\
  ={\eta}_{n}\left(
  \left[f-{\eta}_{n}(f)\right]^2\right)
\displaystyle+
\sum_{1\leq p< n}~(1-{\eta}_{p-1}(g)^2)~\frac{{\eta}_{p-1}(g)}{{\eta}_{n-1}(g)}
{\eta}_{p}\left(\left[Q_{p,n}(f-\eta_n(f))\right]^2\right). 
  \end{array}
  $$

  \subsubsection{Variances at equilibrium}\label{var-equilibrium-sec}
  Starting at equilibrium $\eta_0=\pi$ we have
  \begin{eqnarray*}
E_0~w^{{\tiny hard}}_{n}(f)
&=&E_0^{-1}~{v}^{{\tiny hard}}_{n}\left(
f-\pi(f)\right)\\
&  =&\pi\left(
  \left[f-\pi(f)\right]^2\right)
\displaystyle+\left(1-E_0^2\right)~
\sum_{1\leq p< n}~
\pi\left(\left[\frac{Q^{p}}{E_0^{p}}(f-\pi(f))\right]^2\right) \\
&=&w^{{\tiny soft}}_{n-1}(f)=v^{{\tiny soft}}_{n-1}(f-\pi(f))
 \end{eqnarray*}
 
  and
  $$
  \begin{array}{l}
\displaystyle E_0^{-2} v_{n+1}^{{\tiny hard}}(f)
\displaystyle=\sum_{0\leq p\leq  n} \pi K\left(\left[\frac{1_S}{E_0}~\frac{{Q}^{p}(f)}{
E_0^{p}}-
\pi(f)\right]^2\right)- E_0~\displaystyle\sum_{1\leq p\leq  n}~
\pi\left(\left[
\frac{Q^{p}(f)}{E_0^{p}}-
\pi(f)\right]^2\right).
  \end{array}
  $$
  On the other hand, we have
  \begin{eqnarray*}
   \pi K\left(\left[\frac{1_S}{E_0}~\frac{{Q}^{p}(f)}{
E_0^{p}}-
\pi(f)\right]^2\right)&=& \frac{1}{E_0}~\pi K\left(\frac{1_S}{E_0}~\left[\frac{{Q}^{p}(f)}{
E_0^{p}}\right]^2\right)-\pi(f)^2\\
&=&\frac{1}{E_0}~\pi \left(\left[\frac{{Q}^{p}(f)}{
E_0^{p}}\right]^2\right)-\pi(f)^2\\
&=&\frac{1}{E_0}~\pi \left(\left[\frac{{Q}^{p}(f)}{
E_0^{p}}-\pi(f)\right]^2\right)+\left[\frac{1}{E_0}-1\right]~\pi(f)^2.
  \end{eqnarray*}
  This implies that
   $$
  \begin{array}{l}
\displaystyle E_0^{-1}~ v_{n+1}^{{\tiny hard}}(f)
\displaystyle=~\pi \left(\left[f-\pi(f)\right]^2\right)+(n+1)\left[1-E_0\right]~\pi(f)^2\\
\\\hskip5cm+
\displaystyle\left(1- E^2_0\right)~\displaystyle\sum_{1\leq p\leq  n}~
\pi\left(\left[
\frac{Q^{p}(f)}{E_0^{p}}-
\pi(f)\right]^2\right)\\
\\
\displaystyle=E^2_0~\pi \left(\left[f-\pi(f)\right]^2\right)+(n+1)\left[1-E_0\right]~\pi(f)^2\\
\\\hskip3cm+
\displaystyle\left(1- E^2_0\right)~\displaystyle\left[v_{n}^{{\tiny soft}}(f)+\sum_{1\leq p\leq  n}
\pi\left(\left[~Q\left\{\frac{{Q}^{p-1}(f)}{E_0^{p-1}}-\pi(f)\right\}\right]^2\right) \right]\\
\\
\displaystyle\leq E^2_0~\pi \left(\left[f-\pi(f)\right]^2\right)+(n+1)\left[1-E_0\right]~\pi(f)^2\\
\\\hskip3cm+
\displaystyle\left(1- E^2_0\right)~\displaystyle\left[v_{n}^{{\tiny soft}}(f)+E_0~\sum_{0\leq p<  n}
\pi \left(\left[\frac{{Q}^{p}(f)}{E_0^{p}}-\pi(f)\right]^2\right) \right].\
  \end{array}
  $$
  Choosing $f=\varphi_0$ we find that (\ref{equilibrium-hard-2}).

\subsubsection*{Acknowledgements}
We thank Persi Diaconis for discussions that led to the writing of this article.
AJ was supported by a Singapore Ministry of Education Academic Research Fund Tier 2 grant: R-155-000-161-112.

\appendix

\section*{Appendix}

\subsection*{Proof of the variance formulas (\ref{var-formula-dp}) and  (\ref{var-formula-is})}\label{proof-var-formula-dp}

We have
\begin{eqnarray*}
v_n^{{\tiny dp}}(f)
&=&\EE(\varphi_0^{-1}(Y^h_n))^{-2}~N~\EE\left(\left[\frac{1}{N}\sum_{1\leq i\leq N}\varphi_0^{-1}(Y^{\varphi, i}_n)~f(Y^{\varphi, i}_n)-\EE(\varphi_0^{-1}(Y^{\varphi}_n)~f(Y^{\varphi}_n))
\right]^2\right)\\
&=&\eta_0(\varphi_0)~\frac{E_0^n}{\eta_0Q^n(1)}~\eta_n(f/\varphi_0)-\eta_n(f)^2.
\end{eqnarray*}
In the same vein we have
\begin{eqnarray*}
w_n^{{\tiny dp}}(f)&=&\EE\left(\left[\frac{\varphi_0^{-1}(Y^{\varphi}_n)}{\EE\left(h^{-1}(Y^{\varphi }_n)\right)}(f(Y^{\varphi}_n)-\eta_{n}(f))\right]^2
\right)\\
&=&  \frac{E^n_0}{\eta_0 Q^n(1)}~\eta_0(\varphi_0)~\eta_n\left(\varphi_0^{-1}\left[f-\eta_{n}(f)\right]^2\right)
\leq \rho_Q~\rho(\varphi_0)~\eta_n\left(\left[f-\eta_{n}(f)\right]^2\right).
\end{eqnarray*}
This ends the proof of (\ref{var-formula-dp}). 

Now we come to the proof of (\ref{var-formula-is}). 
We have
\begin{eqnarray*}
v_n^{{\tiny IS}}(f)&:=&N~\EE\left(\left[\frac{\Za_n^{{\tiny IS}}}{\Za_n}~\eta^{{\tiny IS}}_{n}(f)-\eta_{n}(f)
\right]^2\right)\\
&=&\Za_n^{-2}~N~\EE\left(\left[\frac{1}{N}\sum_{1\leq i\leq N}~(Z_n(Y^i)~f(Y^i_n)-\EE(Z_n(Y)~f(Y_n))
\right]^2\right)\\
&=&\EE\left(\left(\frac{Z_n(Y)}{\Za_n}\right)^2~f(Y_n)^2\right)-\eta_n(f)^2=\frac{\eta_0\widetilde{Q}^n(1)}{[\eta_0{Q}^n(1)]^2}~
\widetilde{\eta}_n(f^2)-\eta_n(f)^2.
\end{eqnarray*}
In much the same way, we have
\begin{eqnarray*}
w_n^{{\tiny IS}}(f):=N~\EE\left(\left[\eta_{n}^{{\tiny IS}}(f)-\eta_{n}(f)
\right]^2\right)&=&\EE\left(\left[\frac{Z_n(Y)}{\Za_n}(f-\eta_{n}(f))\right]^2\right)\\
&=&\Za_n^{-2}~\eta_0\widetilde{Q}^n\left(\left[f-\eta_{n}(f)\right]^2\right).
\end{eqnarray*}
Recalling that
\begin{eqnarray*}
\eta_0\widetilde{Q}^n\left[\left[f-\eta_{n}(f)\right]^2\right)&=&\eta_0\widetilde{Q}^n(f^2)-2\eta_n(f)~\eta_0\widetilde{Q}^n(f)+\eta_{n}(f)^2~\eta_0\widetilde{Q}^n(1)\\
&=&\eta_0\widetilde{Q}^n(1)\left[\widetilde{\eta}_n(f^2)-2\eta_n(f)\widetilde{\eta}_n(f)+\eta_n(f)^2\right]\\
&=&\eta_0\widetilde{Q}^n(1)~\left\{\widetilde{\eta}_n\left([f-\widetilde{\eta}_n(f)]^2\right)+\left(\widetilde{\eta}_n(f)-\eta_n(f)\right)^2\right\}
\end{eqnarray*}
we obtain the formula
$$
w_n^{{\tiny IS}}(f)=\frac{\eta_0\widetilde{Q}^n(1)}{[\eta_0{Q}^n(1)]^2}~\left\{\widetilde{\eta}_n\left([f-\widetilde{\eta}_n(f)]^2\right)+\left(\widetilde{\eta}_n(f)-\eta_n(f)\right)^2\right\}.
$$
This ends  the proof of (\ref{var-formula-is}). \cqfd
\subsection*{Mean field particle samplers}\label{mean-field-section}

Notice that
\begin{equation}\label{def-Phi-nl}
\eta_{n+1}=\eta_{n}\Ka_{\eta_n}\quad\mbox{\rm with}\quad \Ka_{\eta_n}(x,z):=(\Sa_{\eta_n}M)(x,z):=\sum_{y\in S}
\Sa_{\eta_n}(x,y)~M(y,z)
\end{equation}
and the Markov transition
$$
\Sa_{\eta_n}(x,y)=g(x)~\delta_x(y)+(1-g(x))~\Psi(\eta_{n})(y).
$$
The mean field particle interpretation of the the measure valued equation (\ref{def-Phi-nl}) is a Markov chain $\xi_n=(\xi^i_n)_{1\leq i\leq N}$ on the product space
$S^N$,
starting with $N$ independent random variables $\xi_0:=(\xi^{i}_0)_{1\leq i\leq N}$ with common law $\eta_{0}$.
The transition of the chain are given for any $ x=(x_i)_{1\leq i\leq N}\in S^N$  by
$$
\PP\left(\xi_{n+1}=x~|~\xi_{n}\right)=\prod_{1\leq i\leq N}~\Ka_{m(\xi_{n})}(\xi^i_n,x_i)
\quad\mbox{\rm
with}\quad
m(\xi_{n}):=\frac{1}{N}\sum_{1\leq i\leq N}\delta_{\xi^{i}_{n}}.
$$
This particle model (a.k.a.~sequential Monte Carlo) is a genetic type particle model with a acceptance-rejection selection transition and an exploration transition dictated by the potential function $g$ and the Markov transition $M$.

For path-space models we also have
\begin{equation}\label{evol-Qa}
\QQ_{n+1}=\Phi_{n+1}(\QQ_{n}):=\Psi_{g_{n}}(\QQ_{n})M_{n+1}
\end{equation}
where $M_{n+1}$ stands for the Markov transition of the historical process $\Xb_n$ and $g_{n}$
is the extension of $g$ to path-spaces defined by
$$
\forall (x_0,\ldots,x_n)\in S_n\qquad
g_{n}(x_0,\ldots,x_n):=g(x_n).
$$

 The mean field particle 
approximation associated with the updating-prediction equation (\ref{evol-Qa}) is defined as above with $N$ path-space valued particles
$$
\chir^i_n=\left(\chir^i_{0,n},\chir^i_{1,n},\ldots,\chir^i_{n,n}\right).
$$ 
 
This mean field particle sampler coincide with the evolution of the genealogical tree of the genetic type
particles defined above. In this interpretation, the path-space particles can be interpreted as the  ancestral lines of the individual at every time step.

By (\ref{def-Phi-hat}) we also have
\begin{equation}\label{def-Phi-nl-hat}
\widehat{\eta}_{n+1}=\widehat{\eta}_{n}\widehat{\Ka}_{\widehat{\eta}_n}\quad\mbox{\rm with}\quad \widehat{\Ka}_{\widehat{\eta}_n}:=\widehat{\Sa}_{\widehat{\eta}_n}K
\end{equation}
and the Markov transition
$$
\widehat{\Sa}_{\widehat{\eta}_n}(x,y)=1_S(x)~\delta_x(y)+(1-1_S(x))~\Psi(\widehat{\eta}_{n})(y).
$$
The mean field particle interpretation of the the measure valued equation (\ref{def-Phi-nl-hat}) is a Markov chain $\widehat{\xi}_n=(\widehat{\xi}^i_n)_{1\leq i\leq N}$ on the product space
$S^N$,
starting with $N$ independent random variables $\widehat{\xi}_0=\xi_0$.  This particle model  is also a genetic type particle model with a acceptance-rejection selection transition and an exploration transition dictated by the potential function $1_A$ and the Markov transition $K$.

\subsection*{Local fluctuation variance formula}
Following the fluctuation analysis developed in chapter 9 in~\cite{delmoral1}, and in chapters 14 and 16 in~\cite{delmoral2}, the limiting variances of the soft particle samplers
presented in section~\ref{soft-obstacle-samplers} are expressed in terms of a sequence of independent centered Gaussian fields $U_{n+1}$ with covariance functions defined by teh formula
  \begin{eqnarray*}
 \EE(U_{n+1}(f)^2)&=&\displaystyle\lim_{N\rightarrow\infty}N~\EE\left(\left[\eta^{{\tiny soft}}_{n+1}(f)-\Phi\left(\eta^{{\tiny soft}}_{n}\right)(f)\right]^2\right)
 \\
 &=&\displaystyle\frac{1}{2}\sum_{x\in S}\eta_n(x)~\Ka_{\eta_n}(x,y)~\Ka_{\eta_n}(x,z)~\left(f(y)-f(z)\right)^2\\
 &=&\eta_{n+1}(f^2)-\eta_n\left(Q(f)^2\right)\\
 &&\hskip3cm
-\eta_{n+1}(f)~\left[\eta_n((1-g)^2)~\eta_{n+1}(f)+2\eta_n\left((1-g)Q(f)\right)\right]
  \end{eqnarray*}
 with the matrix $
  Q(x,y)=g(x)M(x,y)
  $. We check the last assertion using the decompositions
  \begin{eqnarray*}
 \EE(U_{n+1}(f)^2)
 &=&\eta_n\left(g^2\left[M(f^2)-M(f)^2\right]\right)+\eta_n((1-g)^2)~\left[\eta_{n+1}(f^2)-\eta_{n+1}(f)^2\right]\\
 &&\hskip.3cm+\eta_n\left((1-g)Q(f^2)\right)+\eta_n\left(g(1-g)\right)\eta_{n+1}(f^2)-2\eta_n\left((1-g)Q(f)\right)\eta_{n+1}(f).
  \end{eqnarray*}
  For centered functions, the above formulas become
    $$
   \EE(\left[U_{n+1}(f-\eta_{n+1}(f))\right]^2)=\eta_{n+1}(f^2)-\eta_n\left(Q(f)^2\right).
  $$

\subsection*{Proof of the variance formulae (\ref{v-soft-f})}  

By theorems 16.2.1 and 16.6.2~in~\cite{delmoral2} we have
\begin{eqnarray*}
v_n^{{\tiny soft}}(f)&
\displaystyle=&\eta_0\left(\left[\frac{{Q}^{n}(f)}{\eta_0{Q}^{n}(f)}-\eta_n(f)\right]^2\right)+
\sum_{1\leq p\leq  n}\EE\left(\left(~U_{p}
\left[\frac{{Q}^{n-p}(f)}{\eta_{p} {Q}^{n-p}(1)}-\eta_n(f)\right]~\right)^2\right).  \end{eqnarray*}
This implies that
$$
\begin{array}{l}
\displaystyle v_n^{{\tiny soft}}(f)\\
\\
\displaystyle=\sum_{0\leq p\leq   n}\eta_p\left(\left[\frac{{Q}^{n-p}(f)}{\eta_{p} {Q}^{n-p}(1)}-\eta_n(f)\right]^2\right)
-
\displaystyle\sum_{1\leq p\leq   n}~
\eta_{p-1}\left(\left[~Q\left\{\frac{{Q}^{n-p}(f)}{\eta_{p}{Q}^{n-p}(1)}-~\eta_n(f)\right\}\right]^2\right)\\
\\
=\displaystyle\sum_{0\leq p\leq   n}\eta_p\left(\left[\frac{{Q}^{n-p}(f)}{\eta_{p} {Q}^{n-p}(1)}-\eta_n(f)\right]^2\right)
-
\displaystyle\sum_{0\leq p<   n}~\eta_{p}(g)^2~
\eta_{p}\left(\left[~\frac{{Q}^{n-p}(f)}{\eta_{p} {Q}^{n-p}(1)}-\frac{g}{\eta_{p}(g)}~\eta_n(f)\right]^2\right).
 \end{array}
$$
The last assertion follows from
$$
\begin{array}{l}
\displaystyle\eta_{p-1}\left(\left[~Q\left\{\frac{{Q}^{n-p}(f)}{\eta_{p}{Q}^{n-p}(1)}-~\eta_n(f)\right\}\right]^2\right)
=
\displaystyle\eta_{p-1}\left(\left[~\left\{\frac{\eta_{p-1}(g)~{Q}^{n-(p-1)}(f)}{\eta_{p-1}{Q}^{n-(p-1)}(1)}-g~\eta_n(f)\right\}\right]^2\right).\
 \end{array}
$$

By theorem 14.4.3 in~\cite{delmoral2} we also have
$$
  \begin{array}{l}
\displaystyle w_n^{{\tiny soft}}(f)\\
\\
=\displaystyle\eta_0\left(\left[\frac{{Q}^{n}}{\eta_0{Q}^{n}(1)}(f-\eta_n(f))\right]^2\right)+
\sum_{1\leq p\leq  n}\EE\left(\left(U_{p}
\left[\frac{{Q}^{n-p}}{\eta_{p} {Q}^{n-p}(1)}\left(f-\eta_{n}(f)\right)\right]~\right)^2\right)\\
\\
\\=\displaystyle\eta_0\left(\left[\frac{{Q}^{n}}{\eta_0{Q}^{n}(1)}(f-\eta_n(f))\right]^2\right)\\
\\
\displaystyle+
\sum_{1\leq p\leq  n}\left[\eta_p\left(\left[\frac{{Q}^{n-p}}{\eta_p{Q}^{n-p}(1)}(f-\eta_n(f))\right]^2\right)-\eta_{p-1}\left(\left[\frac{{Q}^{n-(p-1)}}{\eta_p{Q}^{n-p}(1)}(f-\eta_n(f))\right]^2\right)\right].
  \end{array}
$$
This implies that
$$
  \begin{array}{l}
\displaystyle\lim_{N\rightarrow\infty}w_n^{{\tiny soft}}(f)\\
\\
=\displaystyle\eta_0\left(\left[\frac{{Q}^{n}}{\eta_0{Q}^{n}(1)}(f-\eta_n(f))\right]^2\right)
\displaystyle+
\sum_{1\leq p\leq  n}\left[\eta_p\left(\left[\frac{{Q}^{n-p}}{\eta_p{Q}^{n-p}(1)}(f-\eta_n(f))\right]^2\right)\right.\\
\\
\displaystyle\hskip4cm\left.-(\eta_{p-1}(g)^2-1+1)~
\eta_{p-1}\left(\left[\frac{{Q}^{n-(p-1)}}{\eta_{p-1}{Q}^{n-(p-1)}(1)}(f-\eta_n(f))\right]^2\right)\right]\\
\\
\displaystyle=\eta_n\left(\left[f-\eta_n(f)\right]^2\right)
\displaystyle+
\sum_{0\leq p<  n}(1-\eta_{p}(g)^2)~
\eta_{p}\left(\left[\frac{{Q}^{n-p}}{\eta_{p}{Q}^{n-p}(1)}(f-\eta_n(f))\right]^2\right).
  \end{array}
$$

%
%

\subsection*{Proof of proposition~\ref{proposition-ratio-functions}}  
Recalling that
\begin{eqnarray*}
\displaystyle\sum_{1\leq l\leq d}\sin{\left(\frac{il\pi}{d+1}\right)}&=&\displaystyle\frac{\cos{\left(\frac{i\pi}{2(d+1)}\right)}
-\cos{\left(\frac{(2d+1)i\pi}{2(d+1)}\right)}}{2\sin{\left(\frac{i\pi}{2(d+1)}\right)}}\\
&=&\displaystyle\frac{\cos{\left(\frac{i\pi}{2(d+1)}\right)}}{2\sin{\left(\frac{i\pi}{2(d+1)}\right)}}~(1-(-1)^i)=\left\{
\begin{array}{ccl}
0&\mbox{\rm if}&~\mbox{\rm $i$ is even}\\
\cot{\left(\frac{i\pi}{2(d+1)}\right)}&\mbox{\rm if}&~\mbox{\rm $i$ is odd}
\end{array}
\right.
\end{eqnarray*}
we find that
\begin{eqnarray*}
Q^{n}(1)(k)&=&\frac{2}{d+1}~\sum_{1\leq i~\mbox{\footnotesize odd} \leq d}~E_{i-1}^n~
\sin{\left(\frac{i}{d+1}~k\pi\right)}~\cot{\left(\frac{i}{d+1}~\frac{\pi}{2}\right)}\\
&=&\frac{4}{d+1}~\sum_{1\leq i~\mbox{\footnotesize odd} \leq d}~E_{i-1}^n~\cos^2{\left(\frac{i}{d+1}~\frac{\pi}{2}\right)}
~U_{k-1}\left(\cos{\left(\frac{i}{d+1}~\pi\right)}\right)~\end{eqnarray*}
with the 
Chebychev polynomial of the second kind
$$
U_k(\cos{\alpha})=\frac{\sin{((k+1)\alpha)}}{\sin{(\alpha)}}\in [-(k+1),(k+1)].
$$
When $k=1$ we have
\begin{eqnarray*}
Q^{n}(1)(1)&=&\frac{4}{d+1}~\sum_{1\leq i\mbox{\tiny odd} \leq d}~E_{i-1}^n~\cos^2{\left(\frac{i}{d+1}~\frac{\pi}{2}\right)}.
\end{eqnarray*}
This readily implies that
$$
Q^{2n}(1)(k)\leq (k+1)~Q^{2n}(1)(1)\Longrightarrow \rho(Q^{2n}(1))\leq (d+1)
$$
and for odd powers 
$$
\frac{Q^{2n+1}(1)(1)}{Q^{2n+1}(1)(k)}=\sum_{x}
\frac{g(1)M(1,x)~}{\sum_{y}g(k)M(k,y)~Q^{2n}(1)(y)/Q^{2n}(1)(x)} \leq (d+1)~\rho(g).
$$

Working a little harder we observe that the extreme value of the function $$1\leq i~\mbox{\footnotesize odd} \leq d\mapsto U_{k-1}\left(\cos{\left(\frac{i}{d+1}~\pi\right)}\right)$$ is attained at $i=1$ and it is given by
$$
U_{k-1}\left(\cos{\left(\frac{1}{d+1}~\pi\right)}\right)=
\frac{\sin{\left(~k~\frac{1}{d+1}~\pi \right)}}{\sin{\left(~\frac{1}{d+1}~\pi \right)}}.
$$
This implies that
$$
Q^{2n}(1)(k)\leq \frac{\sin{\left(~k~\frac{1}{d+1}~\pi \right)}}{\sin{\left(~\frac{1}{d+1}~\pi \right)}}~
Q^{2n}(1)(1)\Longrightarrow~\rho(Q^{2n}(1))\leq  \frac{1}{\sin{\left(~\frac{1}{d+1}~\pi \right)}}.
$$
The end of the proof of the proposition is now easily completed.\cqfd

\subsection*{Proof of theorem~\ref{spectral-comparison-theo}}\label{proof-spectral-comparison-theo}

We set $A=(2+\theta)~R$ and $B=(2+\theta)~Q$. We have
$$
\lambda_{\tiny max}(A)=(2+\theta)~\widetilde{E}_0\quad\mbox{\rm and}\quad
\lambda_{\tiny max}(B)=(2+\theta)~E_0.
$$
This shows that
$$
{\widetilde{E}_0}/{E_0^{2}}=
\frac{\lambda_{\tiny max}(A)}{\lambda_{\tiny max}(B)^2}~(2+\theta)=(2+\theta)~\frac{\lambda_{\tiny max}(A)}{\left(\theta+2\cos\left(\frac{\pi}{(d+1)}\right)\right)^2}.
$$
This yields the equivalence
\begin{equation}\label{equivalence-R-Q}
\begin{array}{l}
{\widetilde{E}_0}/{E_0^{2}}\geq 1+\delta\left(\theta+2\cos\left(\frac{\pi}{(d+1)}\right)\right)^{-2}\\
\\
\Longleftrightarrow  (2+\theta)\lambda_{\tiny max}(A)\geq  
\left(\theta+2\cos\left(\frac{\pi}{(d+1)}\right)\right)^2+\delta.
\end{array}
\end{equation}

On the other hand, the largest eigenvalue of a symmetric matrix $A=(A(i,j))_{1\leq i, j\leq d}$ is given by
$$
\lambda_{\tiny max}(A)=\max_{v\in\SS^{d-1}}v^{\prime}Av
$$
where $\SS^{d-1}=\{v\in\RR^d~:~\sum_{1\leq i\leq d}v_i^2=1\}$ stands for the unit Euclidian sphere. 
Choosing
$$
v(k)=\frac{1}{\sqrt{d}}~\varphi_{0,\theta}(k)=\sqrt{
\frac{2}{d+1}}~
\sin{\left[\frac{k}{d+1}~\pi\right]}
$$
we have
\begin{eqnarray*}
v^{\prime}Av
&=&v^{\prime}Bv+2\left(2\sqrt{\epsilon_{\theta}}-1\right)~v(1)v(2)-(1-\epsilon_{\theta})~\theta~
(v(1)^2+v(2)^2)\\
&&\\
&=&v^{\prime}Bv+2\left(2\sqrt{\epsilon_{\theta}}-1-\theta~(1-\epsilon_{\theta})\right)~v(1)v(2)-(1-\epsilon_{\theta})~\theta~
(v(1)-v(2))^2\\
&&\\
&=&\displaystyle\left(\theta+2\cos{\left(\frac{\pi}{d+1}\right)}\right)\displaystyle+\left(2\sqrt{\frac{1+\theta}{2+\theta}}-1\right)~\frac{4}{d+1}
\sin{\left[\frac{1}{d+1}~\pi\right]}\sin{\left[\frac{2}{d+1}~\pi\right]}\\
&&\hskip3cm\displaystyle-\frac{\theta}{2+\theta}~\frac{2}{d+1}~\left(
\sin^2{\left[\frac{1}{d+1}~\pi\right]}+\sin^2{\left[\frac{2}{d+1}~\pi\right]}\right).
\end{eqnarray*}
Using the fact that
$$
(1\geq )~\left(2\sqrt{\frac{1+\theta}{2+\theta}}-1\right)=\frac{1}{2+\theta}\left(2\sqrt{{(1+\theta)}{(2+\theta)}}-(2+\theta)\right)\geq \frac{\theta}{2+\theta}
$$
we find that

\begin{eqnarray*}
(2+\theta)~v^{\prime}Av
&=&v^{\prime}Bv+2\left(2\sqrt{\epsilon_{\theta}}-1\right)~v(1)v(2)-(1-\epsilon_{\theta})~\theta~
(v(1)^2+v(2)^2)\\
&=&\left(\theta+2\cos{\left(\frac{\pi}{d+1}\right)}\right)^2+2(2+\theta)\left(2\sqrt{\epsilon_{\theta}}-1-\theta~(1-\epsilon_{\theta})\right)~v(1)v(2)\\
&&\hskip2cm+4\cos{\left(\frac{\pi}{d+1}\right)} \left(1-\cos{\left(\frac{\pi}{d+1}\right)}\right)\\
\\
&&\hskip3cm+\theta~\left[2 \left(1-\cos{\left(\frac{\pi}{d+1}\right)}\right)-
(v(1)-v(2))^2\right].
\end{eqnarray*}
Recalling that
\begin{eqnarray*}
\sin{\left[\frac{1}{d+1}~\pi\right]}-\sin{\left[\frac{2}{d+1}~\pi\right]}&=&-2\sin{\left[\frac{1}{2(d+1)}~\pi\right]}\cos{\left[\frac{3}{2(d+1)}~\pi\right]}\\
1-\cos{\left(\frac{\pi}{d+1}\right)}&=&2\sin^2{\left[\frac{1}{2(d+1)}~\pi\right]}\\
\sin{\left[\frac{1}{d+1}~\pi\right]}\sin{\left[\frac{2}{d+1}~\pi\right]}&=&2\sin^2{\left[\frac{1}{d+1}~\pi\right]}\cos{\left[\frac{1}{d+1}~\pi\right]}
\end{eqnarray*}
we check that
\begin{eqnarray*}
(2+\theta)~v^{\prime}Av
&=&\left(\theta+2\cos{\left(\frac{\pi}{d+1}\right)}\right)^2+2(2+\theta)\left(2\sqrt{\epsilon_{\theta}}-1-\theta~(1-\epsilon_{\theta})\right)~v(1)v(2)\\
&&\hskip1cm+4\cos{\left(\frac{\pi}{d+1}\right)} \left(1-\cos{\left(\frac{\pi}{d+1}\right)}\right)\\
\\
&&\hskip2cm+4\theta~\sin^2{\left[\frac{\pi}{2(d+1)}\right]}~\left[\frac{d-1}{d+1}+\frac{2}{d+1}
\sin^2{\left[\frac{3}{2(d+1)}~\pi\right]}
\right].
\end{eqnarray*}
This proves that
$$
(2+\theta)\lambda_{\tiny max}(A)\geq  
\left(\theta+2\cos\left(\frac{\pi}{(d+1)}\right)\right)^2+\delta
$$
with
\begin{eqnarray*}
\delta
&=&2^3\cos{\left(\frac{\pi}{d+1}\right)} \left\{
\frac{2}{d+1}\left(\sqrt{(1+\theta)(2+\theta)}-(1+\theta)\right)~\sin^2{\left[\frac{\pi}{d+1}\right]}
+\sin^2{\left[\frac{\pi}{2(d+1)}\right]}\right\}\\
&&+2^2\theta~\sin^2{\left[\frac{\pi}{2(d+1)}\right]}~\left[\frac{d-1}{d+1}+\frac{2}{d+1}
\sin^2{\left[\frac{3}{2(d+1)}~\pi\right]}
\right]\\
&\geq &2^3\sin^2{\left[\frac{\pi}{2(d+1)}\right]}~\left[\cos{\left(\frac{\pi}{d+1}\right)} \left\{1+
\frac{2}{d+1}\left(\sqrt{(1+\theta)(2+\theta)}-(1+\theta)\right)
\right\}\right.\\
&&\hskip4cm\left.+\theta~\left(\frac{1}{2}-\frac{1}{d+1}
\cos^2{\left[\frac{3}{2(d+1)}~\pi\right]}
\right)\right]\\
&\geq &4\sin^2{\left[\frac{\pi}{2(d+1)}\right]}~\\
&&\hskip1cm\times~\left[\theta~\left(1-\frac{2}{d+1}\right)+2\cos{\left(\frac{\pi}{d+1}\right)} \left\{1+
\frac{2}{d+1}~\frac{\sqrt{1+\theta}}{\sqrt{2+\theta}+\sqrt{1+\theta}}\right\}\right].
\end{eqnarray*}
This implies that
\begin{eqnarray*}
\widetilde{E}_0/E_0^{2}&\geq &1+4\sin^2{\left[\frac{\pi}{2(d+1)}\right]}~\left[\theta+2\cos{\left(\frac{\pi}{d+1}\right)} 
\right]^{-2}\\
&&\hskip1cm\times~\left[\theta~\left(1-\frac{2}{d+1}\right)+2\cos{\left(\frac{\pi}{d+1}\right)} \left\{1+
\frac{2}{d+1}~\frac{\sqrt{1+\theta}}{\sqrt{2+\theta}+\sqrt{1+\theta}}\right\}\right]\\
&=&1+4\sin^2{\left[\frac{\pi}{2(d+1)}\right]}~\left[\theta+2\right]^{-2} E_0^{-2}
\\
&&\hskip1cm\times~\left[\theta~\left(1-\frac{2}{d+1}\right)+2\cos{\left(\frac{\pi}{d+1}\right)} \left\{1+
\frac{2}{d+1}~\frac{\sqrt{1+\theta}}{\sqrt{2+\theta}+\sqrt{1+\theta}}\right\}\right].
\end{eqnarray*}
Equivalently, we have
\begin{eqnarray*}
\widetilde{E}_0&\geq &E_0^{2}+\frac{2}{(\theta+2)^2}~\left(1-\cos{\left(\frac{\pi}{d+1}\right)}\right)~
\\
&&\hskip1cm\times~\left[\theta~\left(1-\frac{2}{d+1}\right)+2\cos{\left(\frac{\pi}{d+1}\right)} \left\{1+
\frac{2}{d+1}~\frac{\sqrt{1+\theta}}{\sqrt{2+\theta}+\sqrt{1+\theta}}\right\}\right].
\end{eqnarray*}

We also have the upper bound
\begin{eqnarray*}
\lambda_{\tiny max}(A)&\leq &\lambda_{\tiny max}(B)+\max_{v\in\SS^{d-1}}{\left(\left(2\sqrt{\epsilon_{\theta}}-1\right)~2v(1)v(2)-(1-\epsilon_{\theta})~\theta~
(v(1)^2+v(2)^2)\right)}\\
&=&\left(\theta+2\cos{\left(\frac{\pi}{d+1}\right)}\right)\\
&&\hskip1cm+\frac{1}{2+\theta}\max_{v\in\SS^{d-1}}{\left(
\left(2\sqrt{(1+\theta)(2+\theta)}-(2+\theta)\right)~2v(1)v(2)-\theta~
(v(1)^2+v(2)^2)\right)}\\
&\leq &\left(\theta+2\cos{\left(\frac{\pi}{d+1}\right)}\right)+\frac{1}{2+\theta}
\left[\left(2\sqrt{(1+\theta)(2+\theta)}-(2+\theta)\right)-\theta\right].
\end{eqnarray*}
This implies that
\begin{eqnarray*}
\widetilde{E}_0/E_0^{2}&\leq& 1+ 2~\frac{\sqrt{\theta+1}}{\sqrt{2+\theta}+\sqrt{\theta+1}}
\left(\theta+2\cos{\left(\frac{\pi}{d+1}\right)}\right)^{-2} +4~\frac{\sin^2{\left(\frac{\pi}{2(d+1)}\right)}}{\theta+2\cos{\left(\frac{\pi}{d+1}\right)}}.
  \end{eqnarray*}
  Equivalently, we have
\begin{eqnarray*}
\widetilde{E}_0&\leq& E_0^{2}+ \frac{2}{\left(\theta+2\right)^{2}}~\left[\frac{\sqrt{\theta+1}}{\sqrt{2+\theta}+\sqrt{\theta+1}}
~ +\left(\theta+2\cos{\left(\frac{\pi}{d+1}\right)}\right)~\left(1-\cos{\left(\frac{\pi}{d+1}\right)}\right)\right].
  \end{eqnarray*}  
Observe that
\begin{eqnarray*}
\frac{\widetilde{E}_0}{E_0^{2}}&\geq &1+~
\displaystyle\frac{\sin^2{\left(\frac{\pi}{2(d+1)}\right)}}{\theta/2+\cos{\left(\frac{\pi}{d+1}\right)}}~\frac{\left[\theta~\frac{d-1}{d+1}+2\cos{\left(\frac{\pi}{d+1}\right)} \left\{1+
\frac{2}{d+1}~\frac{\sqrt{1+\theta}}{\sqrt{2+\theta}+\sqrt{1+\theta}}\right\}\right]}{\theta/2+\cos{\left(\frac{\pi}{d+1}\right)}}\\
&\geq &1+~\frac{2}{3}~
\displaystyle\frac{\sin^2{\left(\frac{\pi}{2(d+1)}\right)}}{\theta/2+\cos{\left(\frac{\pi}{d+1}\right)}}\geq 1+~
\displaystyle\frac{1}{2+\theta}~\frac{4}{3}~\sin^2{\left(\frac{1}{d+1}~\frac{\pi}{2}\right)}\end{eqnarray*}
as soon as $d\geq 2$. 

This ends the proof of the theorem.\cqfd

\subsection*{Some Taylor inequalities}

$$
f(x)=f(y)+(x-y)~f^{\prime}(y)+\frac{(x-y)^2}{2}~f^{\prime\prime}(x)+\frac{(x-y)^3}{3!}~\int_0^1~3(1-t)^2~f^{\prime\prime\prime}(x+t(y-x))~dt.
$$
For any $x\in ]0,\pi/2[$
$$
\begin{array}{ccccl}
\displaystyle x-\frac{x^3}{3!}&\leq& \sin{(x)}&\leq&\displaystyle x-\frac{x^3}{3!}~\cos{(x)}\\
&&&&\\
\displaystyle x+\frac{x^3}{3}&\leq& \tan{(x)}&\leq&\displaystyle  x+\frac{x^3}{3}~\left[1+\tan^2{(x)}~\left(4+3~\tan^2{(x)}\right)\right].
\end{array}
$$
Recalling the formulae
$$
1+u~\leq \frac{1}{1-u}=1+u~\left(1+\frac{u}{1-u}\right)\Leftrightarrow
1-v~\leq \frac{1}{1+v}=1-v~\left(1-\frac{v}{1+v}\right)
$$
which are valid for any $u\not=1$ and $v\not=-1$, 
we also have
\begin{equation}\label{sine-taylor}
\frac{1}{x}~\leq ~\frac{1}{x}~\left(1+\frac{x^2}{3!}~\cos{(x)}\right)
\leq\displaystyle \frac{1}{\sin{(x)}}\leq\displaystyle\frac{1}{x}~\left(1+\frac{x^2}{3!-x^2}\right)\leq
\frac{1}{x}~\left(1+\frac{x^2}{2}\right)\leq \frac{3}{x}
\end{equation}
and
$$
\displaystyle\frac{1}{x}~\left(1-\frac{x^2}{3}~\left[1+\tan^2{(x)}~\left(4+3~\tan^2{(x)}\right)\right]\right)
\leq\displaystyle \cot{(x)}\leq\displaystyle  \displaystyle\frac{1}{x}~\left(1-\frac{x^2}{3+x^2}\right)~\leq~ \frac{1}{x}.
$$

\end{document}